\documentclass[10pt]{article}

\usepackage{amsmath,amssymb,epsfig,color,bbold}
\usepackage{feynmp-auto}
\usepackage[hidelinks]{hyperref}
\usepackage{caption}
\usepackage{tensor}
\usepackage{subcaption}
\usepackage{cancel}
\usepackage{soul}
\usepackage{physics}
\usepackage{slashed}
\usepackage{comment}
\usepackage{psfrag}
\usepackage{graphicx,xspace}
\usepackage{bm}

\usepackage[affil-it, auth-sc]{authblk}
\usepackage{lineno}
\usepackage[sort&compress,numbers]{natbib}
\usepackage{doi}
\usepackage{eso-pic}
\usepackage[capitalise]{cleveref}

\newcommand{\be}{\begin{equation}}  
\newcommand{\ee}{\end{equation}} 
\def\slash#1{#1\!\!\!/\!\,\,}  
\newcommand{\nl}{\nonumber \\ }
\renewcommand{\order}{{\cal O}}
\long\def\symbolfootnote[#1]#2{\begingroup%
\def\thefootnote{\fnsymbol{footnote}}\footnote[#1]{#2}\endgroup}

\def\me{m}
\def\Ee{E}
\def\pe{p}
\def\Em{E_m}
\def\logE{L_E}
\def\logM{L_m}

\newcommand{\iu}{{\rm i}}
\def\e{\mathrm{e}}

\allowdisplaybreaks

\def\arrowang{15}
\def\arrowlen{2.5mm}
\def\decorsize{2.5mm}

\begin{document}

\AddToShipoutPictureFG*{\AtPageUpperLeft{\put(-60,-75){\makebox[\paperwidth][r]{FERMILAB-PUB-25-0773-T}}}}
\AddToShipoutPictureFG*{\AtPageUpperLeft{\put(-60,-60){\makebox[\paperwidth][r]{CALT-TH-2025-032}}}}
\AddToShipoutPictureFG*{\AtPageUpperLeft{\put(-60,-90){\makebox[\paperwidth][r]{CERN-TH-2025-20}}}}

\title{\Large\bf The $Z\alpha^2$ correction to superallowed beta decays \\ 
      in effective field theory and implications for $|V_{ud}|$ }

\author[1]{Zehua Cao}
\author[1,2]{Richard~J.~Hill}
\author[3,4]{Ryan~Plestid}
\author[1,2]{Peter Vander Griend}
\affil[1]{University of Kentucky, Department of Physics and Astronomy, Lexington, KY 40506 USA \vspace{1.2mm}}
\affil[2]{Fermilab, Theoretical Physics Department, Batavia, IL 60510 USA
\vspace{1.2mm}}
\affil[3]{Walter Burke Institute for Theoretical Physics, 
California Institute of Technology, Pasadena, CA 91125 USA\vspace{1.2mm}}
\affil[4]{Theoretical Physics Department, CERN, 1 Esplanade des Particules, CH-1211 Geneva 23, Switzerland\vspace{1.2mm}}

\date{\today}

\maketitle

\begin{abstract}
  \vspace{0.2cm}
  \noindent
  Superallowed ($0^+\rightarrow0^+$) beta decays currently provide the most precise extraction of quark mixing in the Standard Model. Their interpretation as a measurement of $|V_{ud}|$ relies on a reliable first-principles computation of QED radiative corrections expressed as a series in $Z\alpha$ and $\alpha$. In this work, we provide the first model-independent result for two-loop, $O(Z\alpha^2)$, long-distance radiative corrections where the nuclei are treated as heavy point-like particles. 
  We use renormalization group analysis to obtain new results at $O(Z\alpha^3)$ 
  for the coefficient of double-logarithms in the ratio of the maximal beta energy to the inverse nuclear size, $\Em/R^{-1}$.  
  We use the Kinoshita-Lee-Nauenberg theorem to obtain new results at $O(Z^2\alpha^3)$ for the coefficient of logarithms in the ratio of maximal beta energy to the electron mass, $\log(2\Em/\me)$.  
  We identify a structure-dependent, and therefore short-distance, contribution to the traditional $Z\alpha^2$ correction that should be revisited. 
  We provide the first comprehensive update to the long-distance corrections in almost forty years and comment on the impact of our findings for extractions of $|V_{ud}|$. We find that shifts in the long-distance corrections are $2.5\times$ larger than past estimates of their uncertainty, $1.5\times$ larger than the statistical uncertainty from the combined fit of superallowed decays, and about $1/2$ the size of estimated systematic error, which stems dominantly from nuclear structure effects.   
\end{abstract}
\vfill

\begin{fmffile}{feynmffile} 

\fmfset{arrow_ang}{\arrowang}
\fmfset{arrow_len}{\arrowlen}
\fmfset{decor_size}{\decorsize}
    
\fmfcmd{%
vardef middir(expr p,ang) = dir(angle direction length(p)/2 of p + ang) enddef;
style_def arrow_left expr p = shrink(.7); cfill(arrow p shifted(4thick*middir(p,90))); endshrink enddef;
style_def arrow_left_more expr p = shrink(.7); cfill(arrow p shifted(6thick*middir(p,90))); endshrink enddef;
style_def arrow_right expr p = shrink(.7); cfill(arrow p shifted(4thick*middir(p,-90))); endshrink enddef;}

\section{Introduction}

The lifetimes of superallowed ($0^+\rightarrow 0^+$) beta emitters have an immediate connection to fundamental constants of Nature, and in particular $|V_{ud}|$ \cite{Donoghue:1992dd,Hardy:2020qwl}. Due to a number of approximate symmetries and small expansion parameters \cite{Donoghue:1992dd}, $|V_{ud}|$ can be extracted with a high level of precision (currently estimated at $\sim 3 \times 10^{-4}$)~\cite{Hardy:2020qwl}. An important set of inputs for this program are long-distance QED radiative corrections, which are required at high perturbative order. Recently, two of us have found an error in the $O(Z^2\alpha^3)$ correction \cite{Hill:2023acw,Borah:2024ghn}, which results in sizable shifts to $|V_{ud}|$ (e.g., an estimated shift of $3.2 \times 10^{-4}$ in the $\mathcal{F}t$ value for  $^{26m}{\rm Al}$ \cite{Hill:2023acw}). 

It has recently been realized that all long-distance electromagnetic effects can be computed using a heavy particle effective field theory (EFT), including the Fermi function and other radiative corrections~\cite{Hill:2023acw,Hill:2023bfh,Plestid:2024eib,Borah:2024ghn,VanderGriend:2025mdc,Cao:2025lrw}. This 
provides a model independent definition of the corrections (supplanting previous estimates based on the independent particle model \cite{Sirlin:1986cc,Sirlin:1986hpu,Jaus:1986te}), 
and allows the full machinery of renormalization-group improved perturbation theory to be employed.  
For superallowed decays only one composite operator appears at leading power,
\begin{align}\label{eq:superallowedL}
 {\cal L}_{\rm eff} &=  -{\cal C}(\mu)(\phi_v^{[A,Z]})^* \phi_v^{[A,Z+1]} \bar{\nu}_e 
 \slashed{v}
 (1-\gamma_5) e + {\rm H.c.} \,.
\end{align}
For given final states, the matrix element 
\begin{equation}
    \mathcal{M}(\mu)= \left\langle (\phi_v^{[A,Z]})^* \phi_v^{[A,Z+1]} \bar{\nu}_e 
 \slashed{v}
 (1-\gamma_5) e \right \rangle  ~, 
\end{equation} 
can be computed perturbatively using heavy particle Feynman rules \cite{Hill:2023acw,Plestid:2024eib,Borah:2024ghn} (see \cite{Bernard:2004cm,Ando:2004rk,Cirigliano:2022hob,Cirigliano:2023fnz,Cirigliano:2024nfi,VanderGriend:2025mdc,Cao:2025lrw} for related work on neutron beta decay using heavy-baryon theories). 
Physical, and therefore $\mu$-independent, amplitudes are given by the product of $\mathcal{C}(\mu)$ and $\mathcal{M}(\mu)$. 

A complementary program based on both chiral effective theory \cite{Cirigliano:2023fnz,Cirigliano:2024msg,Cirigliano:2024nfi,Cirigliano:2024rfk,King:2025fph} and dispersion relations \cite{Seng:2018yzq,Seng:2018qru,Czarnecki:2019mwq,Seng:2020wjq,Shiells:2020fqp,Seng:2022cnq,Gennari:2024sbn} has been launched examining short-distance modes, and allows for control over the resummation of electroweak logarithms, and nuclear structure dependent effects; lattice QCD can also play an important role at the hadronic scale \cite{Seng:2020wjq,Feng:2020zdc,Ma:2023kfr}. When matching onto \cref{eq:superallowedL}, all such short distance effects (including those stemming from nuclear structure) are subsumed into $\mathcal{C}(\mu)$ at leading power.
At subleading power other Wilson coefficients multiplying higher dimensional operators appear, but are numerically subleading. 

To compute the lifetime of a beta emitter through $O(Z\alpha^2)$ we require corrections to the rate for $A\rightarrow B \nu e$ through $O(Z\alpha^2)$, 
and corrections to the rate for $A\rightarrow B \nu e \gamma$ through relative $O(Z\alpha)$. 
In this work we present two new theoretical results. First, we compute 
relative $O(Z\alpha)$ virtual corrections to the real photon spectrum, 
finding that the 
result factorizes into the product of the first-order Fermi function
and the tree-level real photon spectrum.
Second, we obtain 
the two-loop virtual corrections at $O(Z\alpha^2)$
in the limit of a massless electron by making use of recently obtained results for neutron decay~\cite{VanderGriend:2025mdc}.
These two results are combined with existing analyses of the operator's anomalous dimension to supply a new 
numerical evaluation 
of long-distance QED corrections to nuclear beta decay.
We include terms enhanced by 
logarithms in the ratio of the maximal beta energy to the inverse
nuclear size, $L=\log(\Lambda/\Em)$ with $\Lambda \sim 1/R^{-1}$, up to 
$O(Z^2 \alpha^4 L)$~\cite{Borah:2024ghn}.  
Using our $O(Z\alpha^2)$ 
result, we apply the Kinoshita-Lee-Nauenberg (KLN) theorem \cite{Kinoshita:1962ur,Lee:1964is} to 
further control terms at 
$O(Z^2\alpha^3, Z\alpha^3, \alpha^2)$ 
that are enhanced by 
logarithms in the ratio of maximal beta energy to the electron mass,
$\log(\Em/m)$.  

The rest of the paper is organized as follows: In \cref{VR}, we explain why there are no $O(Z\alpha^2)$ corrections from real emission beyond those captured by the Fermi function. \Cref{VV} presents the extraction of the $O(Z\alpha^2)$ double-virtual correction to $A \rightarrow B e^+ \nu_e$.  In \cref{PhaseSpace}, we combine these results and integrate over phase space to provide the $O(Z\alpha^2)$ radiative correction to the lifetime of a superallowed emitter $A$. We provide an update for the long-distance corrections 
computed using 
\cref{eq:superallowedL}, including an exploration of the numerical impact of some known higher order corrections.  We conclude with a summary and discussion in \cref{Discussion}.

\pagebreak 

\section{$\order(Z\alpha^2)$ correction to the photon spectrum \label{VR} }
In this section, we demonstrate that real-virtual corrections at $O(Z\alpha^2)$ are given by the product of the Fermi function and the Sirlin $g$ function (more precisely its EFT analog).\!\footnote{Real photons were not explicitly discussed in past calculations of the $Z\alpha^2$ correction~\cite{Sirlin:1986cc,Sirlin:1986hpu,Jaus:1986te}. } 
This factorization stems not from a separation of scales, but because most of the virtual corrections are out of phase (i.e., they are ``pure imaginary'') 
with the leading order real emission amplitude and, therefore, do not contribute to the rate at this order; at higher orders, we do not expect this factorization to hold. 

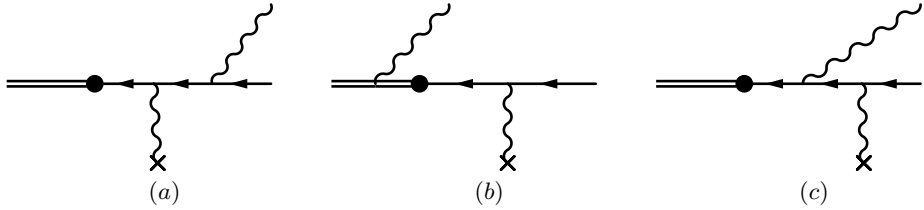
\begin{figure}[t]
\centering
\begin{subfigure}[]{0.25\textwidth}
\vspace{5mm}
\parbox{40mm}{
\begin{fmfgraph*}(100,60)
\fmfstraight
  \fmfleftn{l}{3}
  \fmfrightn{r}{3}
  \fmfbottomn{b}{15}
  \fmf{phantom}{l2,v1,v2,r2}
  \fmffreeze
  \fmf{fermion}{r2,y,z,v1}
  \fmf{double}{l2,v1}
  \fmffreeze
  \fmf{photon}{y,r3}
  \fmf{photon}{z,b9}
  \fmfv{decor.shape=cross}{b9}
\fmfv{decor.shape=circle,decor.filled=full,decor.size=2mm}{v1}
\end{fmfgraph*}
} 
\subcaption{$(a)$}
\end{subfigure} 
\begin{subfigure}[]{0.25\textwidth}
\vspace{5mm}
\parbox{40mm}{
\begin{fmfgraph*}(100,60)
  \fmfleftn{l}{3}
  \fmfrightn{r}{3}
  \fmfbottomn{b}{10}
  \fmftopn{t}{10}
  \fmf{phantom}{l2,v1,v2,r2}
  \fmffreeze
  \fmf{fermion}{r2,x,v1}
  \fmf{double}{l2,y,v1}
  \fmffreeze
  \fmf{photon}{x,b7}
  \fmf{photon}{y,t5}
  \fmfv{decor.shape=cross}{b7}
  \fmfv{decor.shape=circle,decor.filled=full,decor.size=2mm}{v1}
\end{fmfgraph*}
}
\subcaption{$(b)$}
\end{subfigure}
\begin{subfigure}[]{0.25\textwidth}
\vspace{5mm}
\parbox{40mm}{
\begin{fmfgraph*}(100,60)
\fmfstraight
  \fmfleftn{l}{3}
  \fmfrightn{r}{3}
  \fmfbottomn{b}{15}
  \fmf{phantom}{l2,v1,v2,r2}
  \fmffreeze
  \fmf{fermion}{r2,z,y,v1}
  \fmf{double}{l2,v1}
  \fmffreeze
  \fmf{photon}{y,r3}
  \fmf{photon}{z,b12}
  \fmfv{decor.shape=cross}{b12}
  \fmfv{decor.shape=circle,decor.filled=full,decor.size=2mm}{v1}
\end{fmfgraph*}
} 
\subcaption{$(c)$}
\end{subfigure} 

\caption{Feynman diagrams that correct the real-photon amplitude at $O(Z\alpha)$. The circle denotes the weak vertex, and we use the 
Feynman rules discussed in \cite{Borah:2024ghn,Plestid:2024eib} that make $Z$-counting manifest. \label{fig:real-emission} }
\end{figure}

Real photon emission contributes to the lifetime of any $\beta$-emitter beginning at $O(\alpha)$. 
The relative $O(Z\alpha)$ correction to real photon emission then influences the decay rate at $O(Z\alpha^2)$. 
Let us consider 
\begin{equation}
    A \rightarrow B \nu_e e^+(p) \gamma(q)~,
\end{equation}
and write the matrix element as a Dirac structure to be contracted as $\bar{u}_{\nu_e} \slashed{v}(1-\gamma_5)\mathcal{M} v_e$. 
The matrix element is then expanded as 
\begin{equation}
    \mathcal{M}= \mathcal{M}^{(0)} + \mathcal{M}^{(1)} + \ldots ~,     
\end{equation}
The tree level matrix element for single-photon emission is given by the sum of two terms,
\begin{align}
    \mathcal{M}^{(0)}_a &= e \frac{-\slashed{p}-\slashed{q}+m}{2p\cdot q} \slashed{\epsilon}^*~,
    \nl
    \mathcal{M}^{(0)}_b &= -e\frac{\phantom{-}v\cdot\epsilon^*}{-v\cdot q}~, 
\end{align}
where $v_\mu=(1,0,0,0)$ is the four-velocity of the nucleus ($\mathcal{M}^{(0)}_a$ 
and $\mathcal{M}^{(0)}_b$ correspond to diagrams $(a)$ and $(b)$ in 
\cref{fig:real-emission}, without the additional background field attachment).
At one loop order, three diagrams contribute, (cf.~\cref{fig:real-emission})
\begin{align}
       \mathcal{M}^{(1)}_a &= Ze^2 \int {\dd^D L\over (2\pi)^D} \frac{(-1)}{\vb{L}^2} \frac{\slashed{p} + \slashed{q}+\slashed{L}-\me}{2 (\vb{p}+\vb{q})\cdot \vb{L} + \vb{L}^2 - 2p\cdot q -\iu 0} \slashed{v} \mathcal{M}^{(0)}_a \,, 
    \nl
    \mathcal{M}^{(1)}_b &= Ze^2  \int {\dd^D L \over (2\pi)^D} \frac{(-1)}{\vb{L}^2} \frac{\slashed{p}+ \slashed{L}-m}{2 \vb{p}\cdot \vb{L} + \vb{L}^2 -\iu 0} \slashed{v} \mathcal{M}^{(0)}_b \,,
    \nl
     \mathcal{M}^{(1)}_c &= Ze^3 \int {\dd^D L\over (2\pi)^D} \frac{(-1)}{\vb{L}^2} \frac{\slashed{p}+\slashed{q}+\slashed{L}-\me}{2 (\vb{p}+\vb{q})\cdot \vb{L} + \vb{L}^2 - 2p\cdot q -\iu 0} \slashed{\epsilon}^* \frac{\slashed{p}+ \slashed{L}-\me}{2\vb{p}\cdot \vb{L}+\vb{L}^2-\iu 0} \slashed{v}
    \,.
\end{align}
The loop integrations are Euclidean, performed in $D=3-2\epsilon$ dimensions, and the background field Feynman rule enforces $L^0=v\cdot L = 0$.

At $O(Z\alpha^2)$ we need only consider the real part of the interference between $\mathcal{M}^{(0)}$ and $\mathcal{M}^{(1)}$. 
It turns out that due to the structure of the one-loop integrals, almost all contributions to the matrix element are imaginary. This simple observation stems from the fact that three-dimensional Euclidean integrals yield a prefactor of  $1/(-\Ee^2-\iu 0)^{n/2+\epsilon}$ with $n$ an odd integer and $\Ee^2$ a positive number. For $\epsilon\rightarrow0$ this gives $\iu (-1)^{(n-1)/2}/\Ee^n$ such that $\mathcal{M}^{(1)}$ is out-of-phase with $\mathcal{M}^{(0)}$.
Real contributions arise only in the presence of divergences, where
\begin{equation}
\qty(\frac{1}{-\Ee^2-\iu 0})^{n/2+\epsilon} \times \frac{1}{\epsilon}
= (-1)^{(n-1)/2}E^{-n} \left( {\iu\over \epsilon} - 2\iu \log{E} - \pi + \order(\epsilon) \right) \,.
\end{equation}
The product $(\iu) \times \log( -1-\iu0)$ supplies a real part proportional to $\pi$. Indeed, this is the source of the $O(Z\alpha)$ term 
in the expansion of the Fermi function (for a final-state positron),
\begin{align}
\parbox{40mm}{
\raisebox{15pt}{\begin{fmfgraph*}(100,60)
  \fmfleftn{l}{3}
  \fmfrightn{r}{3}
  \fmfbottomn{b}{10}
  \fmftopn{t}{10}
  \fmf{phantom}{l2,v1,v2,r2}
  \fmffreeze
  \fmf{fermion}{r2,x,v1}
  \fmf{double}{l2,v1}
  \fmffreeze
  \fmf{photon}{x,b7}
  \fmfv{decor.shape=cross}{b7}
  \fmfv{decor.shape=circle,decor.filled=full,decor.size=2mm}{v1}
\end{fmfgraph*}
}}
    &= Z e^2 \int {\dd^D L \over (2\pi)^D}  \frac{1}{\vb{L}^2} \frac{-2E_e-\slashed{L}\slashed{v}}{2 \vb{p}\cdot\vb{L} + \vb{L}^2-\iu 0}
    \nonumber
    \\[-20pt]
    &= 
    \frac{i Z\bar{\alpha}}{\beta} \qty(\frac{\mu^2}{- 4\vb{p}^2 - \iu 0})^{\epsilon} \left[ \frac{1}{2\epsilon} 
    +\left(1 + {m\over \Ee} \slashed{v} \right)  \frac{1}{2(1-2\epsilon)} \right]
\nl
&\to -{\pi Z\bar{\alpha} \over 2\beta} \,.
\end{align}
where in the last line we have kept only the real-piece. Notice that this logarithmic contribution is tied to the infrared divergence as $|\vb{L}|\rightarrow 0$.   The same result is obtained by retaining only the IR divergent contribution to the amplitude:  (``$\sim$" denotes equality up to finite, pure imaginary, parts)
\begin{align}
\label{eq:MFsimp}
\parbox{37mm}{
\raisebox{15pt}{\begin{fmfgraph*}(100,60)
  \fmfleftn{l}{3}
  \fmfrightn{r}{3}
  \fmfbottomn{b}{10}
  \fmftopn{t}{10}
  \fmf{phantom}{l2,v1,v2,r2}
  \fmffreeze
  \fmf{fermion}{r2,x,v1}
  \fmf{double}{l2,v1}
  \fmffreeze
  \fmf{photon}{x,b7}
  \fmfv{decor.shape=cross}{b7}
  \fmfv{decor.shape=circle,decor.filled=full,decor.size=2mm}{v1}
\end{fmfgraph*}
}}
    &\sim Z e^2 \int {\dd^D L \over (2\pi)^D}  \frac{1}{\vb{L}^2} \frac{-2E_e}{2 \vb{p}\cdot\vb{L} + \vb{L}^2 -\iu 0}
    = 
    \frac{i Z\bar{\alpha}}{\beta} \qty(\frac{\mu^2}{- 4\vb{p}^2 - \iu 0})^{\epsilon} \left[ \frac{1}{2\epsilon} 
 \right]
\to -{\pi Z\bar{\alpha} \over 2\beta} \,.
\end{align}
Real contributions to $\mathcal{M}^{(1)}$ occur 
only when a $1/\epsilon$ divergence is present.   
Each of the integrals is UV finite, and their IR limits are given by:\!\footnote{
The amplitudes represent hard matching coefficients, and the IR divergences are canceled by corresponding UV divergences in the soft function.}
\begin{align}
      \mathcal{M}^{(1)}_a &\sim 0 \,,
      \nl
    \mathcal{M}^{(1)}_b &\sim \bigg[ Ze^2  \int {\dd^D L \over (2\pi)^D} \frac{1}{\vb{L}^2} \frac{-2E_e}{2 \vb{p}\cdot \vb{L} + \vb{L}^2} \bigg] \mathcal{M}^{(0)}_b \,,
    \nl
     \mathcal{M}^{(1)}_c &\sim 
      \bigg[ Ze^2  \int {\dd^D L \over (2\pi)^D} \frac{1}{\vb{L}^2} \frac{-2E_e}{2 \vb{p}\cdot \vb{L} + \vb{L}^2} \bigg] \mathcal{M}^{(0)}_a \,.
\end{align}
Using the integral (\ref{eq:MFsimp}), we see that 
\begin{align}
    \mathcal{M}^{(1)} &\sim
    \left(  -{\pi Z\bar{\alpha} \over 2\beta}  \right) 
    \left( \mathcal{M}^{(0)}_a + \mathcal{M}^{(0)}_b \right) 
    = \left(  -{\pi Z\bar{\alpha} \over 2\beta}  \right) \mathcal{M}^{(0)} \,,
\end{align}
and hence after performing spin/polarization sums and phase space averages,
\begin{equation}
    \label{main-result-real}
    {\langle |\mathcal{M}_{1\gamma}|^2 \rangle \over \langle |\mathcal{M}^{(0)}_{1\gamma}|^2 \rangle } = 
     1- \frac{\pi Z\alpha}{\beta} + \order(Z^2\alpha^2) = F(Z\alpha,E)+ \order(Z^2\alpha^2) \,. 
\end{equation}
We expect this simple factorization to break at relative $O(Z^2\alpha^2)$. 
The result (\ref{main-result-real}) holds for massive electrons (i.e., for arbitrary values of $m/E$).  
We have checked this result by explicitly computing the relevant master integrals.

\section{$\order(Z\alpha^2)$ corrections without photon emission \label{VV} }
Next, consider the double-virtual $O(Z\alpha^2)$ correction to the decay rate. 
We can extract the result in the limit of a massless electron from our recent work on neutron beta decay \cite{VanderGriend:2025mdc,Cao:2025lrw}.\!\footnote{We have recently been informed of an explicit calculation by O.~Crosas and E.~Mereghetti for finite lepton mass \cite{Crosas:2025xxx}. Their result agrees with our extraction in the small electron-mass limit. } In what follows we neglect terms of $O(\me^2/E_e^2)$.


Consider the matrix element for $[Z+1] \to [Z] e^+ \nu_e$, as a Dirac structure 
to be contracted as
$\bar{u}_{\nu_e} \slashed{v}(1-\gamma_5)\mathcal{M}_{[Z+1]\to [Z]e^+\nu_e} v_e$.   This can be related to the amplitude for the related process 
$[-Z-1] \to [-Z] e^- \bar{\nu_e}$ contracted as 
$\bar{u}_e \mathcal{M}_{[-Z-1] \to [-Z]e^- \bar{\nu}_e} \slash{v}(1+\gamma_5) v_{\nu_e}$.  The relation, to all orders in QED is
\begin{align}\label{eq:CC}
\mathcal{M}_{[Z+1]\to [Z]e^+\nu_e} 
= \mathcal{C} \overline{\mathcal{M}^*}_{[-Z-1] \to [-Z]e^- \bar{\nu}_e}\mathcal{C}^{-1} \,,
\end{align}
where $\overline{\mathcal{M}} = \gamma^0 \mathcal{M}^\dagger \gamma^0$, 
and $\mathcal{C}$ satisfies $\mathcal{C} = \mathcal{C}^{-1}=\mathcal{C}^T = \mathcal{C}^*$, $\mathcal{C} \gamma^\mu \mathcal{C}^{-1} = -(\gamma^\mu)^*$. Using a ``blob'' for all diagrams not involving the background field (which is denoted by $\times$), the amplitude for $[-Z-1] \to [-Z] e^- \bar{\nu_e}$ can be expanded as 
\begin{align}\label{eq:MHdiagrams}
    \vspace{6pt}
 \mathcal{M}_{[-Z-1] \to [-Z]e^- \bar{\nu}_e}  
&=
\quad
\parbox{30mm}{
\begin{fmfgraph*}(100,55)
  \fmfleftn{l}{3}
  \fmfrightn{r}{3}
  \fmfbottomn{b}{5}
  \fmf{phantom}{l2,v,r2}
  \fmffreeze
  \fmf{fermion}{v,r3}
  \fmf{double}{l2,y,v}
  \fmffreeze
\fmfv{decor.shape=circle,decor.filled=shaded,decor.size=5mm}{v}
\end{fmfgraph*} 
}
\nl
&\quad 
+
(-Z)\times \,
\parbox{30mm}{
\begin{fmfgraph*}(100,50)
  \fmfleftn{l}{3}
  \fmfrightn{r}{3}
  \fmfbottomn{b}{5}
  \fmf{phantom}{l2,v,r2}
  \fmffreeze
  \fmf{fermion}{v,x,r3}
  \fmf{double}{l2,v}
  \fmffreeze
  \fmf{photon}{x,b4}
  \fmfv{decor.shape=cross}{b4}
-    \fmfv{decor.shape=circle,decor.filled=full,decor.size=2mm}{v}
\end{fmfgraph*}
}
\quad + \quad 
(-Z)^2 \times \,
\parbox{35mm}{
\begin{fmfgraph*}(100,50)
  \fmfleftn{l}{3}
  \fmfrightn{r}{3}
  \fmfbottomn{b}{7}
  \fmf{phantom}{l2,v,r2}
  \fmf{phantom}{l1,b5,b6,r1}
  \fmffreeze
  \fmf{fermion}{v,y,x,r3}
  \fmf{double}{l2,v}
  \fmffreeze
  \fmf{photon}{x,b6}
  \fmf{photon}{y,b5}
  \fmfv{decor.shape=cross}{b5}
  \fmfv{decor.shape=cross}{b6}
      \fmfv{decor.shape=circle,decor.filled=full,decor.size=2mm}{v}
\end{fmfgraph*}
}  
\\[3mm]
\nonumber
& 
+ 
(-Z)\times
\left(\rule{0cm}{1.4cm}\right.
\parbox{30mm}{
\begin{fmfgraph*}(100,50)
  \fmfleftn{l}{3}
  \fmfrightn{r}{3}
  \fmfbottomn{b}{7}
  \fmf{phantom}{l2,v,r2}
  \fmf{phantom}{l1,b5,b6,r1}
  \fmffreeze
  \fmf{fermion}{v,y,x,r3}
  \fmf{double}{l2,z,v}
  \fmffreeze
  \fmf{photon}{x,b6}
  \fmf{photon,right}{y,z}
  \fmfv{decor.shape=cross}{b6}
      \fmfv{decor.shape=circle,decor.filled=full,decor.size=2mm}{v}
\end{fmfgraph*}
}
\quad + \quad 
\parbox{30mm}{
\begin{fmfgraph*}(100,50)
  \fmfleftn{l}{3}
  \fmfrightn{r}{3}
  \fmfbottomn{b}{7}
  \fmf{phantom}{l2,v,r2}
  \fmf{phantom}{l1,b5,b6,r1}
  \fmffreeze
  \fmf{fermion}{v,y,x,r3}
  \fmf{double}{l2,z,v}
  \fmffreeze
  \fmf{photon,right=0.6}{x,z}
  \fmf{photon}{y,b5}
  \fmfv{decor.shape=cross}{b5}
      \fmfv{decor.shape=circle,decor.filled=full,decor.size=2mm}{v}
\end{fmfgraph*}
}
\quad + \quad 
\parbox{35mm}{
\begin{fmfgraph*}(100,50)
  \fmfleftn{l}{3}
  \fmfrightn{r}{3}
  \fmfbottomn{b}{5}
  \fmf{phantom}{l2,v1,r2}
  \fmffreeze
  \fmf{fermion}{v1,x,r3}
  \fmf{double}{l2,v1}
  \fmffreeze
  \fmf{photon}{x,a1}
  \fmf{photon}{a2,b4}
  \fmf{fermion,left,tension=0.5}{a1,a2,a1}
  \fmfv{decor.shape=cross}{b4}
      \fmfv{decor.shape=circle,decor.filled=full,decor.size=2mm}{v1}
\end{fmfgraph*}
}
\\[6pt] \nonumber
& \quad + \quad
\parbox{35mm}{
\begin{fmfgraph*}(100,50)
\fmfstraight
  \fmfleftn{l}{3}
  \fmfrightn{r}{3}
  \fmfbottomn{b}{5}
  \fmf{phantom}{l2,v1,r2}
  \fmffreeze
  \fmf{fermion}{v1,w,z,y,r3}
  \fmf{double}{l2,v1}
  \fmffreeze
  \fmf{photon}{z,b4}
  \fmf{photon,right}{y,w}
  \fmfv{decor.shape=cross}{b4}
      \fmfv{decor.shape=circle,decor.filled=full,decor.size=2mm}{v1}
\end{fmfgraph*}
}
\quad + \quad
\parbox{35mm}{
\begin{fmfgraph*}(100,50)
    \fmfstraight
  \fmfleftn{l}{3}
  \fmfrightn{r}{3}
  \fmfbottomn{b}{8}
  \fmf{phantom}{l2,v1,r2}
  \fmffreeze
  \fmf{fermion}{v1,w,z,y,r3}
  \fmf{double}{l2,v1}
  \fmffreeze
  \fmf{photon}{y,b7}
  \fmf{photon,right=1.5}{z,w}
  \fmfv{decor.shape=cross}{b7}
      \fmfv{decor.shape=circle,decor.filled=full,decor.size=2mm}{v1}
\end{fmfgraph*}
} \quad 
\left.\rule{0cm}{1.4cm}\right)
+ \dots \,,\\[-12pt]
\nonumber
\end{align}
where the double line denotes a heavy particle with charge $-1$, the solid line with arrow denotes an electron, and the cross denotes insertion of a background field with charge $+1$.   In particular, at $Z=-1$, this expression reduces to the neutron decay amplitude that we have computed in Ref.~\cite{VanderGriend:2025mdc,Cao:2025lrw}.
Solving for the contributions of interest, we have diagrammatically,

\vspace{3pt}

\begin{align} \label{eq:Zalpha2sol}
& \nonumber \\[-25pt]
&
\parbox{33mm}{
\begin{fmfgraph*}(100,47.5)
  \fmfleftn{l}{3}
  \fmfrightn{r}{3}
  \fmfbottomn{b}{7}
  \fmf{phantom}{l2,v,r2}
  \fmf{phantom}{l1,b5,b6,r1}
  \fmffreeze
  \fmf{fermion}{v,y,x,r3}
  \fmf{double}{l2,z,v}
  \fmffreeze
  \fmf{photon}{x,b6}
  \fmf{photon,right}{y,z}
  \fmfv{decor.shape=cross}{b6}
      \fmfv{decor.shape=circle,decor.filled=full,decor.size=2mm}{v}
\end{fmfgraph*}
}
\quad + \quad 
\parbox{30mm}{
\begin{fmfgraph*}(100,47.5)
  \fmfleftn{l}{3}
  \fmfrightn{r}{3}
  \fmfbottomn{b}{7}
  \fmf{phantom}{l2,v,r2}
  \fmf{phantom}{l1,b5,b6,r1}
  \fmffreeze
  \fmf{fermion}{v,y,x,r3}
  \fmf{double}{l2,z,v}
  \fmffreeze
  \fmf{photon,right=0.6}{x,z}
  \fmf{photon}{y,b5}
  \fmfv{decor.shape=cross}{b5}
    \fmfv{decor.shape=circle,decor.filled=full,decor.size=2mm}{v}
\end{fmfgraph*}
}
\quad + \quad   
\dots
\quad
\nl
\nl
&= 
{\cal M}_H - {\cal M}_H|_{v\to -v}
- \left(\rule{0cm}{1.4cm}\right.
\parbox{30mm}{
\begin{fmfgraph*}(100,50)
  \fmfleftn{l}{3}
  \fmfrightn{r}{3}
  \fmfbottomn{b}{5}
  \fmf{phantom}{l2,v,r2}
  \fmffreeze
  \fmf{fermion}{v,x,r3}
  \fmf{double}{l2,v}
  \fmffreeze
  \fmf{photon}{x,b4}
  \fmfv{decor.shape=cross}{b4}    \fmfv{decor.shape=circle,decor.filled=full,decor.size=2mm}{v}
\end{fmfgraph*}
}
\quad + \quad 
\parbox{35mm}{
\begin{fmfgraph*}(100,50)
  \fmfleftn{l}{3}
  \fmfrightn{r}{3}
  \fmfbottomn{b}{7}
  \fmf{phantom}{l2,v,r2}
  \fmf{phantom}{l1,b5,b6,r1}
  \fmffreeze
  \fmf{fermion}{v,y,x,r3}
  \fmf{double}{l2,v}
  \fmffreeze
  \fmf{photon}{x,b6}
  \fmf{photon}{y,b5}
  \fmfv{decor.shape=cross}{b5}
  \fmfv{decor.shape=cross}{b6}
      \fmfv{decor.shape=circle,decor.filled=full,decor.size=2mm}{v}
\end{fmfgraph*}
}  
\quad + \quad \dots 
\left.\rule{0cm}{1.4cm}\right) \,.
\end{align}
In \cref{eq:Zalpha2sol}, the left-hand side (i.e., the first line) denotes all contributions containing at least one background field photon and one non-background field photon.  
The quantity ${\cal M}_H$ is the hard function describing neutron beta decay~\cite{VanderGriend:2025mdc}. 
The final term in parentheses contains terms with only background field photons.  
Diagrams representing wavefunction renormalization have not been displayed, but are included when computing physical amplitudes. 
At one-loop order, the left-hand side of Eq.~(\ref{eq:Zalpha2sol}) vanishes by definition, and the right-hand side is verified to vanish using  Eq.~(8) of Ref.~\cite{VanderGriend:2025mdc} for 
${\cal M}_H - {\cal M}_H|_{v\to -v}$, and Eq.~(37)  of Ref.~\cite{Hill:2023bfh} for 
the one-loop background field diagram.  
At two-loop order, we compute in the small-$\me$ limit, 
\begin{align}\label{eq:MHdiff}
    {\cal M}_H - {\cal M}_H|_{v\to -v}
    &= {\bar{\alpha}\over 4\pi}\left( F_H^{(1)} - \tilde{F}_H^{(1)}  \right)
+ \left( \bar{\alpha} \over 4\pi \right)^2 \left[ 
 (F_R^{(2)} - \tilde{F}_R^{(2)}) + (F_H^{(2)} - \tilde{F}_H^{(2)}) 
 + F_J^{(1)} ( F_H^{(1)} - \tilde{F}_H^{(1)} ) 
\right]  \,,
\end{align}
where $\tilde{F} = F|_{v\to -v}$ and $F_{H}$ is the hard function, $F_J$ the jet function, and $F_R$ the remainder function  for neutron decay defined in Eq.~(16) of Ref.~\cite{VanderGriend:2025mdc}.  
The background field diagrams in this limit are
\begin{align}\label{eq:bfdiags}
    \parbox{30mm}{
\begin{fmfgraph*}(100,50)
  \fmfleftn{l}{3}
  \fmfrightn{r}{3}
  \fmfbottomn{b}{5}
  \fmf{phantom}{l2,v,r2}
  \fmffreeze
  \fmf{fermion}{v,x,r3}
  \fmf{double}{l2,v}
  \fmffreeze
  \fmf{photon}{x,b4}
  \fmfv{decor.shape=cross}{b4}    \fmfv{decor.shape=circle,decor.filled=full,decor.size=2mm}{v}
\end{fmfgraph*}
}
~~~&= {\bar{\alpha}} \left[ {\iu\over 2} \left(2L_E - \iu {\pi} -1 \right)  \right]
\,,
\nl
\nl
\parbox{35mm}{
\begin{fmfgraph*}(100,50)
  \fmfleftn{l}{3}
  \fmfrightn{r}{3}
  \fmfbottomn{b}{7}
  \fmf{phantom}{l2,v,r2}
  \fmf{phantom}{l1,b5,b6,r1}
  \fmffreeze
  \fmf{fermion}{v,y,x,r3}
  \fmf{double}{l2,v}
  \fmffreeze
  \fmf{photon}{x,b6}
  \fmf{photon}{y,b5}
  \fmfv{decor.shape=cross}{b5}
  \fmfv{decor.shape=cross}{b6}
      \fmfv{decor.shape=circle,decor.filled=full,decor.size=2mm}{v}
\end{fmfgraph*}
}  
~&=
 \bar{\alpha}^2 \left[ -\frac18 (2L_E - \iu \pi )^2 + \frac54 - {\pi^2\over 12} \right] \,,
\end{align}
where $\logE=\log(2\Ee/\mu)$. 
Using Eqs.~(\ref{eq:MHdiff}), (\ref{eq:bfdiags}) in 
Eq.~(\ref{eq:Zalpha2sol}) thus determines the $\order(Z\alpha^2)$ contributions 
to the nuclear amplitude in Eqs.~(\ref{eq:CC}) and (\ref{eq:MHdiagrams}).  
Together with the $Z\alpha^2$ contribution including real photon emission 
from Eq.~(\ref{main-result-real}), and the known $\order(\alpha)$ correction, 
the total electron spectrum for $[Z+1]\to [Z] e^+\nu_e$ is 
\begin{equation}
    \label{main-result-msbar}
    \frac{\dd \Gamma}{\dd \Ee}
    = \left(\frac{\dd \Gamma}{\dd \Ee}\right)_{\rm tree} \!\!\!\!\!\! \times\!
    \tilde{F}(Z\bar{\alpha}_1,\Ee,\mu)\! \times \! \qty(1+ \frac{\bar{\alpha}_1}{2\pi} \qty(\hat{g}(E,\mu)+\frac{11}{4}) + Z\bar{\alpha}_1^2 \qty[  - \qty(\frac{2n_e}{3}+1)\logE + \frac{8n_e}{9}+\frac{11}{4} -\zeta_2] + \ldots )  \,,  
\end{equation}
where $\tilde{F}(Z\bar{\alpha}_1,E)$ is the all-orders expression containing the leading $Z$ dependence, cf. Eq.~(\ref{FF-def}) below,\!\footnote{The definition (\ref{FF-def}) includes the additional factor $4\eta/(1+\eta)^2 = \order( (Z\alpha)^4 )$ appearing in the complete $\overline{\rm MS}$ hard function~\cite{Hill:2023acw}.
}
and $\hat{g}(\Ee,\mu)$ is the Sirlin $g$-function but with $m_p$ replaced by the $\overline{\rm MS}$ subtraction point $\mu$ (cf. Eq.~(\ref{eq:sirling}) below; the term $11/4$ converts between $\overline{\rm MS}$ and Sirlin's scheme~\cite{Hill:2023acw}).
Here $n_e$ is the number of massless fermions (i.e., electrons) in the theory; in practice we will always take $n_e=1$. We have expressed the result in terms of the $\overline{\rm MS}$ coupling with $n_e=1$ electron flavor. 
This is related to onshell $\alpha$ (or equivalently $\bar{\alpha}_0$) by $\alpha = \bar{\alpha}_1 \left[1 - \tfrac{n_e}{3\pi} \bar{\alpha}_1 \logM + (\tfrac{n_e}{3\pi} \bar{\alpha}_1 \logM)^2+ \ldots \right]$, where $\logM=\log(m^2/\mu^2)$.
We remark that
the $Z\bar{\alpha}_1^2$ correction remains finite as $\me\rightarrow0$. This finite $\me\rightarrow0$ limit is guaranteed by the KLN theorem \cite{Kinoshita:1962ur,Lee:1964is} (see \cref{app:KLN} and related discussion in \cite{Sirlin:1986cc}) and provides a non-trivial check of our result.

Conventionally, the $O(Z\alpha^2)$ correction is defined via a multiplicative factor,\!\footnote{Our notation tracks powers of $Z$ and $\alpha$ separately. The notation used by Sirlin and Zucchini is related by $\delta_2(\Ee)=\delta^{(2,1)}(\Ee)$ and $\delta_3=\delta^{(3,2)}(\Ee)$. }
\begin{equation}
    \frac{\dd \Gamma}{\dd \Ee} = \left(\frac{\dd \Gamma}{\dd \Ee}\right)_{\rm tree}\!\!\!\!\!\! \times \tilde{F}(Z \alpha,\Ee, \mu) \!\times\!  \qty(1+ \delta^{(1,0)}(\Ee) +  \delta^{(2,1)}(\Ee) + \ldots )  ~,  
\end{equation}
with the series defined in terms of onshell $\alpha$. 
Translating from Eq.~(\ref{main-result-msbar}), we have
\begin{equation}
    \delta^{(1,0)}(\Ee) = 
    \frac{\alpha}{2\pi} \qty(\hat{g}(\Ee,\mu) + \frac{11}{4})~,
\end{equation}
\begin{equation}
    \label{main-result-VV}
    \delta^{(2,1)}(\Ee) = Z\alpha^2 \qty[ \frac{n_e}{3}  \logM -\qty(\frac{2n_e}{3} +1)\logE+\frac{8 n_e}{9}+\frac{11}{4}-\zeta_2~ + O\qty(\frac{\me^2}{\Ee^2})]~. 
\end{equation}
We note that the energy dependence and the dependence on the electron mass match the existing calculation by Sirlin and Zucchini \cite{Sirlin:1986cc}. The finite contribution differs, as expected, since our calculation is performed in a (model-independent) point-like EFT as opposed to the independent particle model used in \cite{Sirlin:1986cc}.

Using our observation that real-virtual corrections factorize at 
$O(Z\alpha^2)$, the relation (\ref{eq:Zalpha2sol}) for the $\order(Z\alpha^2)$ virtual corrections, and the result from Ref.~\cite{VanderGriend:2025mdc} for the two-loop correction at $Z=-1$, we thus obtain the new result (\ref{main-result-VV}) for $\delta^{(2,1)}(\Ee)$, representing the complete result for the long-distance QED corrections through $O(Z\alpha^2)$ in the limit of a massless electron.  
Our recent work on neutron beta decay also supplies the virtual corrections at $O(\alpha^2)$ in the limit of a massless electron, but $O(\alpha^2)$ real-virtual and real-real corrections are currently unknown (as are the finite parts of the $O(Z^2\alpha^3)$ corrections).

\subsection*{Comment on short-distance terms}
The correction derived in \cref{main-result-VV} is computed in the low-energy effective theory where nuclear structure is not explicitly present. In past literature it has been emphasized that nuclear structure modifies the resulting correction \cite{Sirlin:1986cc,Sirlin:1986hpu,Jaus:1986te}, and so before proceeding we comment on how these effects appear in the EFT formalism. 

First, logarithms of the form $\log(\Lambda/2\Em)$ (where $\Lambda$ is a nuclear scale) are captured by RG evolution, and specifically the $O(Z\alpha^2)$ coefficient of the anomalous dimension; these terms are included in our results. Second, there can  be finite corrections induced at the nuclear scale. These are absorbed as $O(Z\alpha^2)$ corrections to the Wilson coefficient $\mathcal{C}(\mu)$ and depend on nuclear structure. 

In the context of the present paper, it is sufficient to show that these corrections factorize from the long-distance matrix element. This is guaranteed because they are absorbed inside the Wilson coefficient $\mathcal{C}(\mu)$ which appears as the tree-level vertex at the heart of every diagram (at arbitrary order in perturbation theory). They are therefore not a part of the long-distance corrections that we discuss below, but instead arise from short-distance regions that probe the scale of nuclear structure. They should be properly booked as a $O(Z\alpha^2)$ contribution to the nuclear-structure dependent corrections (sometimes called $\delta_{\rm NS}$ in the literature). 

For the present work, we have taken an ansatz for the short distance correction from the work of Sirlin and Zucchini as implemented by Towner and Hardy \cite{Hardy:2020qwl,Towner:2007np}. After subtracting the $\overline{\rm MS}$ low-energy matrix element this may be interpreted as their estimate for the structure-dependent short-distance region. We write this as 
$\mathcal{C}(\Lambda) \to \mathcal{C}(\Lambda)\left(1 +  \frac{1}{2}\delta_{\rm NS,TH}^{(2,1)}\right)$, 
where $\delta_{\rm NS,TH}^{(2,1)}$ is defined in \cref{app:relating-schemes}.

\section{Correction to the lifetime \label{PhaseSpace} }
Let us now combine the above results and obtain a prediction for the decay rate. We are interested in long-distance QED corrections that are typically defined after extracting the leading result containing powers of $Z\alpha$. 
In the EFT approach we follow here this is given by \cite{Hill:2023acw,Hill:2023bfh} (all expressions are for positron emitters)
\begin{equation}    \label{FF-def}
   \tilde{F}(Z \alpha ,E,\mu) =  \frac{4 \eta}{(1+\eta)^2} \frac{2 (1+\eta)}{[\Gamma(2\eta +1)]^2} |\Gamma(\eta +\iu\xi)|^2 \e^{\pi \xi} (2 p r)^{2(\eta-1)}~,
\end{equation}
where $r= 1/(\e^{\gamma_{\rm E}}\mu)$ (with $\gamma_{\rm E}$ the Euler-Mascheroni constant), and $\eta=\sqrt{1-Z^2\alpha^2}$, $\xi= -Z\alpha \Ee/\pe$.
We define 
\begin{align}    
    \label{outer-def}
    1+\delta_R'(\mu_H) &\equiv   R(\mu_H,\mu_L) M(\mu_L,\Em,\me) + O\qty(\frac{q_{\rm ext}}{\Lambda})~,
\end{align}
with (notice that $1+\delta_R'(\mu_H)$ is independent of $\mu_L$),
\begin{align}
    \label{outer-def-R}
    R(\mu_H,\mu_L)&=\qty[ \frac{\mathcal{C}(\mu_L)\big/\mathcal{C}(\mu_H)}{\exp\qty[(1-\eta)~\log(\mu_H/\mu_L)]}]^2~,\\
    \label{outer-def-M}
    M(\mu_L,\Ee,\me)&= \frac{\int \dd \Pi \, |\mathcal{M}(\mu_L)|^2 }{\int \dd \Pi ~ \tilde{F}(Z \alpha,E,\mu_L) }~,
\end{align}
where $\Em \sim 10~{\rm MeV}$ is set by the nuclear mass splitting, $\Lambda \sim 100~{\rm MeV}$ represents a short distance scale, and $\int\dd \Pi$ denotes integration over electron, neutrino, and photon phase space (when considering real emission).  We do not consider terms suppressed by $q_{\rm ext}/\Lambda$, where $q_{\rm ext}$ denotes the  external lepton momentum, which arise from higher dimensional operators in the point-like effective theory. 

We have split the long-distance correction $(1+\delta_R')$ into a piece induced by renormalization group (RG) flow, $R(\mu_H,\mu_L)$, and the low-energy matrix element contribution $M(\mu_L,\Em,\me)$. The denominators are the full amplitude in the limit that $\alpha\rightarrow 0$ with $Z\alpha$ held fixed.  The RG term is governed by the anomalous dimension and QED beta function, contains all large logarithms, and does not depend on energy. The matrix-element term is evaluated at a low-scale $\mu_L \sim q_{\rm ext}$ and does not contain large logarithms.  With the above definition, we may write the decay rate as  
\begin{equation}
    \Gamma =   \qty( \frac{1}{\pi^3} \int_\me^{\Em} \dd \Ee ~\pe \Ee (\Em-\Ee)^2 \tilde{F}(Z,E,\mu_H) ) \times |\mathcal{C}(\mu_H)|^2 \times (1+\delta_R'(\mu_H))   + O\qty(\frac{q_{\rm ext}}{\Lambda})~,
\end{equation}
where $\Em=M_A-M_B=Q_{\rm EC}-\me$
is the energy emitted in the beta decay 
(we neglect nuclear recoil). 
Comparing to the typical parameterization, written as $(1+\Delta_R)(1+\delta_{\rm NS}-\delta_C)(1+\delta_R')$, we see that $\mathcal{C}(\mu_H)$ has subsumed all short distance corrections (including nuclear structure dependence) as well as fundamental constants $G_F$ and $|V_{ud}|$. 

The factor $R(\mu_H,\mu_L)$ accounts for RG evolution from $\mu_H\sim \Lambda$ down to the dynamical scale of the beta decay $\mu_L\sim \Em$. It can be computed to high accuracy and controls contributions that are enhanced by logarithms of the form $\log(\mu_H/\mu_L)\sim\log(\Em R)$ (see \cref{app:running} for details). It is conventional in the beta decay literature to include the Fermi function in the phase space measure. We implement this 
as
\begin{equation} \label{eq:Fav}
    \langle \ldots \rangle_F =  \frac{\int_{\me}^{\Em} \dd \Ee  ~\pe \Ee( \Em-\Ee)^2 \tilde{F}(Z,\Ee,\mu) \times ( \ldots )}{\int_{\me}^{\Em} \dd \Ee  ~\pe \Ee( \Em-\Ee)^2 \tilde{F}(Z,\Ee,\mu) }~.
\end{equation}
The $\mu$ dependence in $\tilde{F}(Z,\Ee,\mu)$ cancels to all orders in $Z\alpha$ 
between numerator and denominator in Eq.~(\ref{eq:Fav}).

\subsection*{Comment on one-loop corrections \label{app:one-loop-comment} }
In this work we restrict ourself to an analysis at leading power, which is equivalent to taking $E R \rightarrow 0$ (and by extension $m R \rightarrow 0$ and $|\vb{p}| R \rightarrow 0$). At subleading power, it is well known that one encounters structure dependent terms that can depend on lepton energy. For example, as we discuss in \cref{app:2008-reproduce}, if one includes certain structure dependent terms in the shape factor or definition of the Fermi function, this can slightly alter the phase-space average of the one-loop radiative corrections. 

The appropriate quantity to use at leading power is the point-like Fermi function or equivalently (for the purposes of phase space averaging) the resummation of all leading-in-$Z\alpha$ diagrams in the EFT. Any structure dependence that depends on $E$ or $\vb{p}$ will correspond to higher dimensional operators in the point-like effective theory, beyond the single operator listed in \cref{eq:superallowedL}. The one-loop renormalization of these operators, and their contribution to the real photon spectrum, is not guaranteed to be captured by the ansatz of $\dd \Phi F(Z,E)  \rightarrow \dd \Phi F(Z,E)  \times (1+\delta^{(1,0)}(E) )$.

In \cref{app:2008-reproduce}, we find that in order to reproduce Towner and Hardy's phase space averages of the one-loop corrections through three digits, structure dependent corrections must be included. We account for  these effects using the analytic expressions taken from Ref.~\cite{Hayen:2017pwg} (see \cref{app:2008-reproduce} for a discussion). The numerical size of 
\begin{equation}
   \frac{\alpha}{2\pi}\langle g(E,\Em,m) \rangle_F  - \frac{\alpha}{2\pi}  \bar{g},  
\end{equation}
with $\bar{g}$ taken from Table V of \cite{Towner:2007np}, gives a sense for the error induced by the factorization ansatz $\dd \Phi F(Z,E) \times (1+\delta^{(1,0)}(E) )$ at sub-leading power. We estimate the uncertainty due to this effect in \cref{app:2008-reproduce}.

\subsection{New results for outer corrections}

\begin{table}[t]
    \centering
    \begin{tabular}{c c c  c  | c c c}
        Beta emitter & $Z_{\rm daughter}$ & $\Lambda~~[{\rm MeV}]$ &  $2 \Em~[{\rm MeV}]$ & $R(\mu_H,\mu_L)$  &  $M(\mu_L)$ &  $1+\delta_R'(\mu_H)$\\
        \hline
        \hline \\[-12pt]
         $^{10}{\rm C}$ & 5 & 199.1 & 2.794 & 1.01634 & 0.99793 & 1.01424(5) \\
        $^{14}{\rm O}$ & 7 & 188.9 & 4.641 & 1.01464 & 0.99805 & 1.01266(3) \\
        $^{26m}{\rm Al}$ & 12 & 159.3 & 7.443 & 1.01309 & 0.99826 & 1.01133(1) \\
        $^{34}{\rm Cl}$ & 16 & 147.2 & 9.961 & 1.01226 & 0.99842 & 1.01066(2) \\
        $^{34}{\rm Ar}$ & 17 & 142.6 & 11.102 & 1.01181 & 0.99844 & 1.01023(2) \\
        $^{38m}{\rm K}$ & 18 & 142.0 & 11.066 & 1.01199 & 0.99851 & 1.01048(3) \\
        $^{42}{\rm Sc}$ & 20 & 137.8 & 11.831 & 1.01191 & 0.99860 & 1.01050(4) \\
        $^{46}{\rm V}$ & 22 & 134.0 & 13.083 & 1.01166 & 0.99869 & 1.01033(6) \\
        $^{50}{\rm Mn}$ & 24 & 132.1 & 14.247 & 1.01152 & 0.99878 & 1.01029(7) \\
        $^{54}{\rm Co}$ & 26 & 130.9 & 15.467 & 1.01141 & 0.99888 & 1.01027(9)
    \end{tabular}
\caption{Radiative corrections computed using the conventions described in the main text for the ten nuclei with the lowest $\mathcal{F}t$ errors in \cite{Hardy:2020qwl}. We fix $\mu_H = \Lambda = \sqrt{6}/\langle r^2\rangle^{1/2}$ with the root-mean-square radius of the daughter, $\langle r^2 \rangle^{1/2}$,  taken from \cite{Angeli:2013epw} (except for $^{34}{\rm Cl}$ for which we use $\langle r^2\rangle^{1/2}=3.39~{\rm fm}$ \cite{Hardy:2004id}) to match \cite{Hardy:2020qwl}, while scale variation errors show dependence on different choices of $\mu_L$.  We take a central value of $\mu_L= 2\Em$ and vary $\mu_L \rightarrow 2 \mu_L$ and $\mu_L \rightarrow \tfrac12 \mu_L$ (for $^{26m}{\rm Al}$ there is an accidental cancellation and we inflate the error estimate by hand to $10^{-5}$). The RG factor is computed using \cref{R-expl-ms-bar} while the matrix element factor is computed using the contributions under the integral in \cref{factorized-form-Gamma-complete}. 
\label{tab:decay_corrections}}
\end{table}

Here we present our results for the point-like long-distance corrections, $(1+\delta_R'(\mu_H))$ defined in this work. We present results for $R(\mu_H,\mu_L)$ and $M(\mu_L)$ separately, and use the residual dependence on $\mu_L$ as an estimate of neglected higher-order perturbative corrections.   
We fix $\mu_H = \Lambda=\sqrt{6}/\langle r^2 \rangle^{1/2}$, where  $\langle r^2 \rangle^{1/2}$ is the root-mean-square radius of the daughter nucleus, and vary $\mu_L = \Em \dots 4 E_m$
We use the form of the expressions in terms of $\bar{\alpha}_L$ given in \cref{R-expl-ms-bar,factorized-form-Gamma-complete}, with central value $\mu_L=2\Em$.

\subsection*{Finite mass corrections}
The scale variation used in \cref{tab:conversion formula} is an estimate of missing finite terms at $O(Z^2\alpha^3)$, i.e., those without a logarithmic enhancement, and other higher order effects. It does not account for any error incurred from the small-mass expansion of $\delta^{(2,1)}(E)$, nor does it account for the point-like approximation used herein. 

To estimate these missing uncertainties, we simply assign an error of $Z\alpha^2 \me^2/E_\me^2$ for finite-mass corrections to $\delta^{(2,1)}(E)$. The size of $\me^2/E_\me^2$ is largest for the lightest nuclei $^{10}{\rm C}$, $^{14}{\rm O}$, and $^{26m}{\rm Al}$; however, the modest values of $Z$ for these nuclei help keep this error under control:
\begin{equation}
    ^{10}{\rm C}:Z\alpha^2 \qty(\frac{\me}{\Em})^2=0.36 \times10^{-4} 
    ~,~
    ^{14}{\rm O}:Z\alpha^2 \qty(\frac{\me}{\Em})^2=0.18 \times10^{-4} 
    ~,~
    ^{26m}{\rm Al}: Z\alpha^2 \qty(\frac{\me}{\Em})^2=0.12 \times10^{-4} ~.
\end{equation}
This quantity is smaller than $1\times10^{-5}$ for all other nuclei.  
The estimated error is thus smaller than the perturbative error estimated in \cref{tab:decay_corrections}, except for the
three lightest beta emitters
where the two uncertainties are comparable.

\subsection*{Mixing of structure-dependent and radiative corrections}
Structure dependent terms that depend on the electron's energy at tree-level can be identified with higher dimensional operators in the effective theory used here.  These operators receive different radiative corrections than the leading order operator in \cref{eq:superallowedL}. To the best of our knowledge, these corrections have not been calculated.  We estimate, on the basis of the results in \cref{app:2008-reproduce}, that these terms are $\sim 10^{-5}$ for light nuclei but can grow to become $\sim 10^{-4}$ for larger-$Z$ beta emitters.  These are smaller than the perturbative error estimated in \cref{tab:decay_corrections}. 

\begin{table}[]
    \centering
    \begin{tabular}{c c|c c c || c}
        Beta emitter & $Z_{\rm daughter}$ & Convention shift  &  Long-distance shift &  Total-shift & $\delta_{\rm NS,TH}^{(2,1)}~~[10^{-4}]$\\
        \hline
        \hline \\[-12pt]
        $^{10}{\rm C}$ & 5 & 1.000105\!\!\! & 0.999967\!\!\! & 1.00007 & 2.1 \\
        $^{14}{\rm O}$ & 7 & 1.00012 & 1.00000 & 1.00012 & 2.9  \\
        $^{26m}{\rm Al}$ & 12 & 1.00013 & 1.00006 & 1.00019 & 4.3 \\
        $^{34}{\rm Cl}$ & 16 & 1.00013 & 1.00014 & 1.00027 & 5.5 \\
        $^{34}{\rm Ar}$ & 17 & 1.00013 & 1.00016 & 1.00029 & 5.8 \\
        $^{38m}{\rm K}$ & 18 & 1.00015 & 1.00017 & 1.00032 & 6.2 \\
        $^{42}{\rm Sc}$ & 20 & 1.000154\!\!\! & 1.000214\!\!\! & 1.00037 & 6.8 \\
        $^{46}{\rm V}$ & 22 & 1.00015 & 1.00027 & 1.00042 &  7.2 \\
        $^{50}{\rm Mn}$ & 24 & 1.00015 & 1.00033 & 1.00048 & 7.6 \\
        $^{54}{\rm Co}$ & 26 & 1.00016 & 1.00039 & 1.00055 & 8.0 
    \end{tabular}
    \caption{Update to the long-distance radiative corrections from Towner and Hardy using the procedure described in \cref{app:relating-schemes} for the ten nuclei considered in this work.  Additional digits are shown when they affect the final digit displayed in the ``Total-shift'' column. The convention shift stems from a reorganization of Towner and Hardy's  calculation of $(1+\delta_R')_{\rm TH}$. The total shift includes both the effects of this rearrangement and the incorporation of the higher order terms calculated in this work. The long-distance shift does not alter the $Z\alpha^2$ short-distance correction, $\delta_{\rm NS,TH}^{(2,1)}$, whose size we estimate using Towner and Hardy's calculation. This quantity is sizeable and should be revisited (in addition to uncertainties on $\delta_{\rm NS}-\delta_{C}$) in a consistent microscopic theory of the nucleus. } 
    \label{tab:conversion formula}
\end{table}

\subsection{Comparison to Towner and Hardy}
In order to take the above analysis and infer the associated shift in $|V_{ud}|$, we reinterpret the analysis of Towner and Hardy~\cite{Hardy:2020qwl}. The details are discussed in \cref{app:relating-schemes},
and we simply quote the main results here. Towner and Hardy's calculation of $\delta_2^{({\rm TH})}$ contains the sum of the short- and long-distance contributions. We do not adjust their value of $\delta_2^{({\rm TH})}$, and rather interpret it as a long-distance piece that precisely matches our formulae, and a short-distance piece that we then infer from their calculation. This short distance piece is interpreted as a $Z\alpha^2$ shift to $\mathcal{C}(\mu)$, or equivalently to an $O(Z\alpha^2)$ contribution to the nuclear-structure dependent correction which we denote as $\delta_{\rm NS}^{(2,1)}$. 

We first reorganize the existing ingredients of Towner and Hardy's calculation according to the factorization theorem that can be derived from \cref{eq:superallowedL} (see \cite{Hill:2023acw} for details). This leads to the shift denoted ``convention shift'' in \cref{tab:conversion formula}. Next we compare the results of the present paper with the long-distance corrections of Towner and Hardy defined in this new convention. This leads to the quantity denoted ``long-distance shift''. The product of ``convention shift" and ``long-distance shift" gives the total shift. 

\begin{figure}
    \centering
    \includegraphics[width=0.666\linewidth]{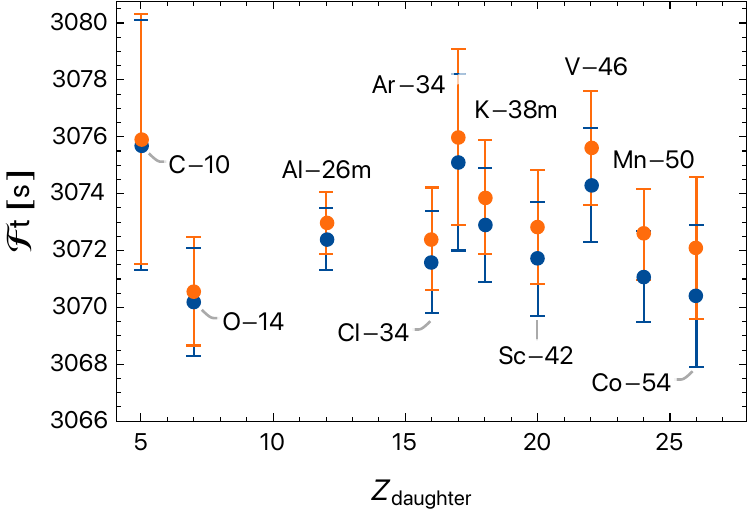}\hspace{0.075\linewidth}
    \caption{A reproduction of Fig.~3 of Towner and Hardy \cite{Hardy:2020qwl} for the $\mathcal{F}t$ values of the ten nuclei considered in this work. Error bars show purely experimental errors, and do not account for theoretical uncertainties related to radiative corrections. The data points of Towner and Hardy are shown in blue (dark gray). The same data points multiplied by the ``total shift'' column of \cref{tab:conversion formula} are shown in orange (light gray). 
    \label{fig:shift-effect}}
\end{figure}

In \cref{fig:shift-effect}, we plot the total shifts from \cref{tab:conversion formula} as applied to the $\mathcal{F}t$ values from Fig.~3 of \cite{Hardy:2020qwl}; we do not adjust the error bars.  The shifts are sizeable. Our analysis incorporates many terms that are not present in Towner and Hardy's analysis.  Let us list these for illustration: {\it i)} sub-leading logarithms at $O(\bar{\alpha}^2 L)$ from the RG evolution [+0.4], {\it ii)} the coefficient at $Z^2\bar{\alpha}^3 L$ [-0.7] and  $Z^2\bar{\alpha}^3 L \logM$ [+1.9], {\it iii)} the coefficient at $O(Z\bar{\alpha}^3 L^2)$ [+0.3], {\it iv)} the coefficient at $O(Z^2 \bar{\alpha}^4 L^3)$ [+0.02], and {\it v)} the coefficient of $\log(\me)$ singularity at $O(Z^2\bar{\alpha}^3)$ [-1.0]. The numbers in square brackets give the shift to the total radiative correction, $1+\delta_R'(\mu_H=\Lambda)$, for $^{26m}{\rm Al}$ (in units of $10^{-4}$) relative to a ``baseline'' where this term is set to zero by hand. 

 If we perform a weighted average of the $\mathcal{F}t$ values of the ten nuclei\footnote{Towner and Hardy take the fifteen most precise superallowed emitters and obtain $\langle \mathcal{F} t\rangle = (3072.24 \pm 0.57)~{\rm s}$ \cite{Hardy:2020qwl}. The $0.57$ error is statistical and should be compared to the estimated theory errors from $\delta_{\rm NS}$ [$\pm 1.72$] and $\delta_R'$ [$\pm 0.36$]. } considered here we find $\langle \mathcal{F} t\rangle = (3072.1 \pm 0.6)~{\rm s}$. If the updated radiative corrections are applied using the shift discussed above, we instead obtain $\langle \mathcal{F} t\rangle = (3073.0 \pm 0.6)~{\rm s}$. This shift is $2.5\times$ larger than the estimated uncertainty from $\delta_R'$ in \cite{Hardy:2020qwl}, is $1.5\times$ larger than the statistical error,  and is roughly half the estimated error from $\delta_{\rm NS}$; the shift exacerbates the first-row CKM unitarity deficit (however see e.g., \cite{Gorchtein:2025wli} for nuclear structure effects which do the opposite).

\section{Summary and discussion \label{Discussion} }
In this work, we have provided a comprehensive update to the long-distance radiative corrections for superallowed beta decays. Our results are derived using a rigorous factorization of the decay amplitude at leading-power (i.e., neglecting the interference of structure-dependent higher dimensional operators and $O(\alpha)$ radiative corrections).  This factorization of the amplitude has allowed us to quantify errors incurred by Towner and Hardy's factorization ansatz, as well as to control higher-order radiative corrections proportional to logarithms of the form $\log(\Lambda/(2\Em))$ and $\log(2\Em/\me)$. We replace Sirlin's heuristic estimate for the $Z^2\alpha^3$ correction with a systematic determination of logarithmic enhancements using renormalization group methods.

Our three main theoretical result are:
\begin{itemize}
    \item The amplitude factorizes at leading-power, and this factorization theorem constrains the possible form of higher order radiative corrections. 
    
    \item The virtual corrections to real-photon radiation  are captured by the Fermi function at $O(Z\alpha)$. 
  
    \item The $O(Z\alpha^2)$ matrix element of the point-like effective theory has been presented. Combined with the results from  \cite{Hill:2023bfh,Borah:2024ghn,Hill:2023acw}, this allows us to control all effects of $O(Z\alpha^2)$ including logarithmic enhancements which we count numerically as $\log(\Lambda/2\Em)\sim \log(2\Em/\me)\sim Z$. 
\end{itemize}
The relevant formulae for these results are given by 
\cref{outer-def,outer-def-M,outer-def-R} for the factorization theorem,
\cref{main-result-real} for real-photon corrections,  and  \cref{main-result-msbar,R-expl-ms-bar} (or \cref{main-result-VV,R-expl} in terms of onshell $\alpha$) for the analytic formulae for long-distance radiative corrections.  Our scale variation uncertainty suggests that currently unknown  $Z^2\alpha^3$ corrections and finite mass effects are $\sim 3 \times 10^{-5}$ in size. This is much smaller than $\delta_{\rm NS, TH}^{(2,1)}$, and this quantity should be revisited in light of its sizeable impact on $|V_{ud}|$.

Future work will involve a careful merger of the long-distance EFT calculations presented here with short-distance results. Ideally, both theories will be formulated using minimal subtraction or another closely related scheme. The results presented here are sufficient for $\sim 10^{-4}$ precision, which we expect to be sufficient for the near-term and medium-term future of precision beta decay experiments.

Motivated by this goal, and given that the long-distance region is perturbatively calculable, we list some open questions in the low-energy theory whose answers would reduce theoretical uncertainty. These include: finite electron-mass corrections to the long-distance $Z\alpha^2$ results presented here; the proper one-loop renormalization of higher dimensional operators; a calculation of the $Z^2\alpha^3$ matrix element in the small electron-mass limit; and an analysis of $O(\alpha^2)$ corrections including real-virtual corrections and double-real emission. Most pressing, however, is the control of the $Z\alpha^2$ contribution to the short-distance matching, $\delta^{(2,1)}_{\rm NS}$, and the control of one-loop radiative corrections sensitive to nuclear structure. We look forward to future progress on these open questions and towards a consistent implementation of the results presented above with a short-distance matching calculation. 

\section*{Acknowledgments}
RP thank Eric Laenen for helpful discussions, and Roberto Zucchini for correspondences relating to Ref.~\cite{Sirlin:1986cc}. 
RJH thanks Susan Gardner for discussions. 
RP thanks the Ohio State University, the University of Illinois Urbana Champaign, the Jo\v zef Stefan Institute, the Kavli Institute for Theoretical Physics, and the Facility for Rare Isotopes and Beams for their hospitality. KITP is supported by the National Science Foundation under Grant No. NSF PHY-1748958. FRIB is a DOE Office of Science User Facility under Award Number DE-SC0023633 
RP was supported for large portions of this work by the Neutrino Theory Network under Award Number DEAC02-07CH11359, the U.S. Department of Energy, Office of Science, Office of High Energy Physics under Award Number DE-SC0011632, and by the Walter Burke Institute for Theoretical Physics. PVG acknowledges support from the Visiting Scholars Award Program of the Universities Research Association. This work was supported by the U.S. Department of Energy, Office of Science, Office of High Energy Physics, under Award DE-SC0019095. This work was produced by Fermi Forward Discovery Group, LLC under Contract No. 89243024CSC000002 with the U.S. Department of Energy, Office of Science, Office of High Energy Physics.

\vspace{24pt}

\textbf{Note added:} After the completion of this work the authors of Ref.~\cite{Crosas:2025xxx} informed us of their own independent calculation of the $Z\alpha^2$ correction. Their analysis includes full electron mass dependence (going beyond the result derived herein), and furthermore includes corrections to correlation coefficients. Our work provides an independent extraction of the small-$\me$ limit of the $Z\alpha^2$ correction, a novel analysis of real-virtual corrections which explains their factorization, higher order renormalization group evolution, and supplies KLN constraints on the amplitude at three-loop order. 

\appendix

\section{Resummation of logarithms 
\label{app:running} }
For convenience we provide here a discussion of the renormalization group flow (see also Ref.~\cite{Hill:2023acw}). The renormalization group equation governing the flow of $\mathcal{C}(\mu)$ is
\begin{equation}
   \frac{\dd \log \mathcal{C}}{\dd \log \mu}  =  \mathcal{\gamma}_{\mathcal{O}}~. 
\end{equation}
Since $\bar{\alpha}(\mu)$ flows with $\mu$, it is convenient to express the solution to this equation as 
\begin{equation}
    \log \qty( \frac{\mathcal{C}(\mu_H)}{\mathcal{C}(\mu_L)}   )  = \int_{\bar{\alpha}_L}^{\bar{\alpha}_H} \frac{\gamma_{\mathcal{O}}(\bar{\alpha})}{~\beta(\bar{\alpha})} \dd \bar{\alpha} ~,
\end{equation}
where the $\beta$ function is defined via 
\begin{equation}\label{eq:beta}
 \beta(\bar{\alpha}) \equiv   \dv{\log \mu} \bar{\alpha}(\mu) = -2\alpha \qty[\frac{\bar{\alpha}}{4\pi} \beta_0  + \qty(\frac{\bar{\alpha}}{4\pi})^2 \beta_1 + \ldots~]~. 
\end{equation}
For QED\footnote{In principle, we could also include a dynamical muon for $\mu \geq m_\mu$ and dynamical pions for $\mu \geq m_\pi$. For simplicity, we opt instead to work exclusively in $n_e=1$ QED. The price of this choice is that matching from a microscopic theory at the scale $\mu_H=\Lambda$ will involve small-logarithms at $O(\alpha^2)$ and $O(Z\alpha^3)$ of the form $\log(\Lambda/m_\mu)\approx 0.5$, $\log(m_\pi/m_\mu)\approx 0.25$ and $\log(\Lambda/m_\pi)\approx0.35$ with numerical values obtained using $\Lambda=190~{\rm MeV}$; these count as $O(\lambda^4)$ (see discussion above \cref{eq:alpha_H}) in our counting and are absorbed into the matching calculation that defines the short-distance correction.} 
the relevant values are $\beta_0=-(4/3) n_e$ and $\beta_1=-4 n_e$. 
One expands the anomalous dimension as 
\begin{equation}
    \begin{split}
    \gamma_{\cal O} = \frac{\bar{\alpha}}{4\pi} &\qty[
    \gamma_0^{(1)} ]
    + \qty(\frac{\bar{\alpha}}{4\pi} )^2 \qty[ Z^2\gamma_1^{(0)} + Z \gamma_0^{(1)} + \gamma_0^{(2)}]
    + \qty(\frac{\bar{\alpha}}{4\pi} )^3 \qty[ Z^2 \gamma_2^{(1)} + \ldots ]+ \qty(\frac{\bar{\alpha}}{4\pi} )^4 \qty[ Z^4\gamma_3^{(0)} + 
    \ldots ]  + \ldots~,
    \end{split}
\end{equation}
where we have retained only the terms needed for the present analysis, 
which can be found in Ref.~\cite{Borah:2024ghn}.

Next, introduce the parameter $\lambda$ and count $\alpha\sim \lambda^2 $ and $Z \sim 1/\lambda$. The RG equations may be expanded and integrated order-by-order in $\lambda$. This yields an expression for the ratio $\mathcal{C}(\mu_H)/\mathcal{C}(\mu_L)$ as a function of $\bar{\alpha}_H\equiv \bar{\alpha}(\mu_H)$ and $\bar{\alpha}_L\equiv \bar{\alpha}(\mu_L)$ as given explicitly in \cite{Hill:2023acw}. 
Integrating Eq.~(\ref{eq:beta}) we obtain
\begin{equation}    
       \label{eq:alpha_H}
    {\bar{\alpha}_H \over \bar{\alpha}_L}
    = 1- \frac{\bar{\alpha}_L}{2\pi} \beta_0 L + 
    \qty(\frac{\bar{\alpha}_L}{2\pi})^2 
    \left(\beta_0^2 L^2 -{\beta_1 \over 2}L \right) +  
    \order(\lambda^4) \,, 
\end{equation}
where we assign the scaling $L=\log(\mu_H/\mu_L)\sim 1/\lambda$. 
The $\overline{\rm MS}$ and onshell couplings are given by\footnote{The largest corrections in our counting are all $O(\lambda)$, e.g., $O(Z \alpha)$ and $Z^2\alpha^2 L$. Therefore, when relating $\alpha$ and $\bar{\alpha}$ one need only work through $O(\lambda^2)$ when a global error of $O(\lambda^3)$ is desired. } 
\begin{equation}
     {\bar{\alpha}(\mu_L)} = \alpha \left[
     1-\frac{\alpha}{2\pi} \beta_0 \log\qty(\frac{\mu_L}{\me}) + \qty(\frac{\alpha}{2\pi})^2 \beta_0^2 \log^2\qty(\frac{\mu_L}{\me}) + \order(\lambda^3) \right] \,,
\end{equation}
where we take $\log(\mu_L/\me)\lesssim 1/\lambda$.\!\footnote{Higher order terms can be obtained from Eqs.~(29)-(32) of  \cite{Baikov:2012rr}.}
Introducing $\logM= \log(\me^2/\mu_L^2)$,
we obtain the renormalization factor in \cref{outer-def-R},
expressed in terms of onshell $\alpha$, 
\begin{equation}
    \label{R-expl}
    \begin{split}
    R(\mu_H,\mu_L) = 1 &+ \alpha \bigg[ {3\over 2\pi} L \bigg]
    + \alpha^2 \bigg[ Z L +  {13 \over 8\pi^2} L^2 + \left(\frac13 - {35\over 48\pi^2}\right) L  - \frac{1}{2\pi^2} L \logM  \bigg]\\
    &+    \alpha^3 \bigg[ {2\over 3\pi}Z^2 L^2 
    + \left( {\pi \over 6} - {3\over \pi}  \right) Z^2 L 
    + {13\over 6\pi} Z L^2 
    + {221 \over 144 \pi^3}L^3
    \\
    &\hspace{0.2\linewidth}
    -\qty(\frac{2 Z(Z+1)}{3 \pi }L  + \frac{13 L^2}{12 \pi ^3}) \logM  + \frac{L}{6\pi^3}L_\me^2 \bigg]
    \\
    &+ \alpha^4 \bigg[ {13\over 9\pi^2} Z^2 L^3 -\frac{5  Z^2}{3 \pi ^2} L^2 \logM + \frac{Z^2}{3 \pi ^2} L L_\me^2 \bigg]  + \order(\lambda^4) \, ~,
    \end{split}
\end{equation}
where we have allowed $\logM \sim 1/\lambda$ (given that $2\Em/m$ is often numerically comparable to $\Lambda/2\Em$ for the nuclei of interest) and truncated our expansion at $O(\lambda^3)$.  \Cref{R-expl} generalizes the equivalent expression given in the Supplementary Material of Ref.~\cite{Hill:2023acw} for $\logM \neq 0$. We also supply the result for $R(\mu_H,\mu_L)$ in terms of $\bar{\alpha}_L$, 
\begin{equation}
    \label{R-expl-ms-bar}
    \begin{split}
    R(\mu_H,\mu_L) = 1 &+ \bar{\alpha}_L \bigg[ {3\over 2\pi} L \bigg]
    + \bar{\alpha}_L^2 \bigg[ Z L +  {13 \over 8\pi^2} L^2 + \left(\frac13 - {35\over 48\pi^2}\right) L \bigg]\\
    &+    \bar{\alpha}_L^3 \bigg[ {2\over 3\pi}Z^2 L^2
    + \left( {\pi \over 6} - {3\over \pi}  \right) Z^2 L 
    + {13\over 6\pi} Z L^2 
    + {221 \over 144 \pi^3}L^3 -{2\over 3\pi}Z^2 L \logM \bigg]
    \\
    &+ \bar{\alpha}_L^4 \bigg[ {13\over 9\pi^2} Z^2 L^3 - \frac{Z^2}{\pi^2} L^2 \logM  - \frac{Z^2}{3\pi^2} L L_\me^2 \bigg]  + \order(\lambda^4) \, ~.
    \end{split}
\end{equation}
Note that the explicit dependence on $\logM$ arises from the conventional definition of $R$ involving onshell $\alpha$.

The analysis written above can be readily extended to $O(\lambda^4)$ with currently available ingredients; this pushes well beyond the $O(10^{-4})$ precision target of superallowed beta decays. Therefore, logarithmic enhancements are completely controlled for the purposes of $|V_{ud}|$ extractions.

\section{Kinoshita-Lee-Nauenberg constraints \label{app:KLN} }
The KLN theorem \cite{Kinoshita:1962ur,Lee:1964is} guarantees that the inclusive decay rate for a superallowed beta decay has a finite limit as $\me\rightarrow 0$. This implies that corrections of the form $\bar{\alpha}_1^2(\mu) \log(\mu/\me)$ (for example)  are forbidden for some fixed $\mu$ as $\me \rightarrow 0$. As a result, we can guarantee that our perturbative expressions are free of any electron-mass logarithms to all-orders in perturbation theory if the result is expressed in terms of $\bar{\alpha}_L$ rather than $\alpha$. This proves to be a powerful constraint on the amplitude as was already appreciated by Sirlin and Zuchinni in their work at $O(Z\alpha^2)$ \cite{Sirlin:1986cc}.

Let us consider the full amplitude 
\begin{equation}
    \Gamma = |\mathcal{C}(\mu_H)|^2 \left| \frac{\mathcal{C}(\mu_L)}{\mathcal{C}(\mu_H)}\right|^2 \int \dd \Phi_e |\mathcal{M}_{0\gamma}(\mu_L)|^2 \qty[ 1+  \int \dd\Phi_{1\gamma}\frac{|\mathcal{M}_{1\gamma}(\mu_L)|^2}{|\mathcal{M}_{0\gamma}(\mu_L)|^2}  + \ldots ] ~,
\end{equation}
where 
$\dd\Phi_e = pE(\Em-E)^2 \dd E$, while $\dd\Phi_{1\gamma}$ is the additional phase space of a single photon defined at fixed electron energy. 

We know the full leading-in-$Z\alpha$ expression for $|\mathcal{M}_{0\gamma}(\mu_L)|^2$ \cite{Hill:2023bfh}, and can readily express the result in terms of $\bar{\alpha}_L$. We may then define corrections as a series in $\bar{\alpha}_L$, 
\begin{equation}
    \begin{split}
    \label{factorized-form-Gamma}
    \Gamma = |\mathcal{C}(\mu_H)|^2 \left| \frac{\mathcal{C}(\mu_L)}{\mathcal{C}(\mu_H)}\right|^2 \int \dd \Phi_e \tilde{F}(Z \bar{\alpha}_L,E,\mu_L)  \bigg[1 &+\frac{\bar{\alpha}_L}{2\pi} \qty(\hat{g}(E,\mu_L)+\frac{11}{4}) + Z\bar{\alpha}_L^2\qty(-\frac{5}{3}\logE+\frac{8}{9}+\frac{11}{4}-\zeta_2) \\
    &\hspace{0.185\linewidth}+ Z^2\bar{\alpha}_L^3 f^{(3,2)} + \bar{\alpha}_L^2 f^{(2,0)} + \ldots ~\bigg]~, 
    \end{split}
\end{equation}
where the functions $f^{(2,3)}$ are currently undetermined, but whose $\log(\me)$ terms will be constrained by the KLN theorem. 
The function $\hat{g}(E,\mu)$ contains electron mass singularities (see e.g., \cite{Sirlin:2011wg}),
\begin{equation} 
    \begin{split}
    \hat{g}(\Ee,\mu) 
    + \frac{11}{4}=~~  &\log\qty(\frac{2\Em}{\me})\qty[ \frac{4(1-x)}{3x} -3 + \frac{(1-x)^2}{6x^2} + 4 \log\qty(\frac{1-x}{x}) ] \\
    &+ 3 \log\qty(\frac{\mu}{2\Em}) +2 - \frac{2\pi^2}{3} + 4 (\log x -1) \qty[ \frac{1-x}{3x}-\frac{3}{2} + \log\qty(\frac{1-x}{x})]\\
    &+ \frac{(1-x)^2}{6x^2}\log x + O\qty(\frac{\me^2}{E^2} \log\frac{\me}{\Ee})~, 
    \end{split}
\end{equation}
where $x=\Ee/\Em$. The electron mass singularity cancels (as is guaranteed by the KLN theorem) when integrated over tree-level phase space, 
\begin{equation}
    \label{key-KLN-identity-1}
     \int_0^1 \dd x ~x^2(1-x)^2 \qty[ \frac{4(1-x)}{3x} -3 + \frac{(1-x)^2}{6x^2} + 4 \log\qty(\frac{1-x}{x}) ] = 0 ~.
\end{equation}
This observation, and the fact that the Fermi function becomes energy independent at $O(Z\alpha)$ for a massless electron, fixes the $\log(\me)$ dependence of $\delta^{(2,1)}$ as was first noticed by Sirlin and Zucchini \cite{Sirlin:1986cc}. 

We may now play the same game to fix the $\log(\me)$ dependence of the $O(\alpha^2)$ and $O(Z^2\alpha^3)$ terms. This is most easily phrased in terms of $\bar{\alpha}_L$ because this quantity is fixed as $m\rightarrow 0$. We demand that $\Gamma$ be given by 
\begin{equation}
    \label{summed-form-Gamma}
    \Gamma = \Gamma^{(0)} + \frac{\bar{\alpha}_L}{2\pi} \Gamma^{(1)}  + \qty(\frac{\bar{\alpha}_L}{2\pi})^2 \Gamma^{(2)}  + \qty(\frac{\bar{\alpha}_L}{2\pi})^3 \Gamma^{(3)}  + \ldots,
\end{equation}
where $\Gamma^{(n)}$ may contain powers of $Z^m$ with $m\leq n$ and logarithmic enhancements of the form $\log(\Lambda/2 \Em)$ but, crucially, {\it does not contain} any terms of the form $\log(2\Em/ m)$ or $\log(\Lambda/m)$.  

Let us compare to the factorized form of \cref{factorized-form-Gamma} prior to integration over phase space. Since $\hat{g}(E,\mu)$ contains a term  proportional to $\log(\me)$, its cancellation in \cref{summed-form-Gamma} is non-trivial. At $O(\bar{\alpha}_L)$, the cancellation proceeds via \cref{key-KLN-identity-1}. At $O(Z\bar{\alpha}_L^2)$, in the small electron mass limit, the Fermi function  reads $\tilde{F}(Z \bar{\alpha}_L,E,\mu_L)= (1-\pi Z \bar{\alpha}_L + \ldots)$, and \cref{key-KLN-identity-1} thus implies that the  $\log(\me)$ cross term between $\tilde{F}(Z \bar{\alpha}_L,E,\mu_L)$ and $\hat{g}(E,\mu)$ again vanishes. A corollary of this observation is that the function appearing at $O(Z\bar{\alpha}_L^2)$ in \cref{summed-form-Gamma} is free of any $\log(\me)$ singularity, in agreement with explicit computation. 

Let us now turn to the undetermined functions $f^{(3,2)}$ and $f^{(2,0)}$. We may immediately conclude that $f^{(2,0)}$ contains no $\log(\me)$ singularities $f^{(2,0)}=0+\ldots$ with the ellipsis denoting $O(1)$ terms as $m\rightarrow 0$. For $f^{(3,2)}$, we instead find that a $\log(\me)$ singularity must be present to cancel against the associated singularity generated by the cross-term between the $Z^2\bar{\alpha}_L^2$ term in the Fermi function, and the $\log(\me)$ singularity from  $\frac{\bar{\alpha}_L}{2\pi} \hat{g}(E,\mu)$. This arises because of the logarithmic energy dependence from
\begin{equation}
    \tilde{F}(Z \bar{\alpha}_L,E,\mu_L)= 1-\pi Z \bar{\alpha}_L  + Z^2\bar{\alpha}_L^2 \qty(\frac{\pi^2}{3}+ \frac{11}{4} - \log\frac{2E}{\mu}) + \ldots ~, 
\end{equation}
leads to the non-vanishing integral
\begin{equation}
    \label{key-KLN-identity-2}
     \int_0^1 \dd x ~x^2(1-x)^2 \qty[ \frac{4(1-x)}{3x} -3 + \frac{(1-x)^2}{6x^2} + 4 \log\qty(\frac{1-x}{x}) ]\times\qty[- \log(x) ]
     = \frac{\pi ^2}{45} -\frac{1}{12}~.
\end{equation}
which implies that\footnote{Strictly speaking we can only conclude that the integral of  $f^{(3,2)}$ over phase space yields $1/12 -\pi^2/45$, however for the purposes of computing the superallowed lifetimes these two statements are equivalent. } (recall the tree-level phase space gives $\int_0^1 \dd x  ~x^2(1-x)^2=1/30$)
\begin{equation}
    \begin{split}
    f^{(3,2)} 
    &= \frac{30}{2\pi} \log\qty(\frac{2\Em}{\me})\qty(\frac{1}{12}-\frac{\pi ^2}{45}) + \ldots \\
    &=  \frac{1}{4\pi}\log\qty(\frac{\me^2}{\mu^2})\qty(\frac{2\pi ^2}{3}-\frac{5}{2}) + \ldots~,   
    \end{split}
\end{equation}
where the ellipses represent terms that are $O(1)$ in the $m\rightarrow 0$ limit, including $\log(2\Em/\mu)$ (hence why the two forms written above are equivalent). We have thus predicted the coefficient of $\log(\me)$ in $f^{(2,3)}$  and $f^{(2,0)}$ (which vanishes) as introduced in \cref{factorized-form-Gamma} on the basis of the KLN theorem. 

Counting $\log(2\Em/m) \sim \log(\Lambda/2\Em) \sim Z \sim 1/\lambda$ and $\bar{\alpha}_L\sim \lambda^2$ as above, we have
\begin{equation}
    \begin{split}
    \label{factorized-form-Gamma-complete}
    \Gamma = |\mathcal{C}(\mu_H)|^2 \left| \frac{\mathcal{C}(\mu_L)}{\mathcal{C}(\mu_H)}\right|^2 \int \dd \Phi_e \tilde{F}(Z \bar{\alpha}_L,E,\mu_L)  \bigg[1 &+\frac{\bar{\alpha}_L}{2\pi} \qty(\hat{g}(E,\mu)+\frac{11}{4}) + Z\bar{\alpha}_L^2\qty(-\frac{5}{3}\logE+\frac{8}{9}+\frac{11}{4}-\zeta_2) \\
    &\hspace{0.15\linewidth} + \frac{1}{4\pi}Z^2\bar{\alpha}_L^3 \logM\qty(\frac{2\pi ^2}{3}-\frac{5}{2})  + O(\lambda^4) \bigg]~.
    \end{split}
\end{equation}
Since we know $\bar{F}(Z \bar{\alpha}_L,E,\mu_L)$ to all-orders in perturbation theory and $\left| \mathcal{C}(\mu_L)/\mathcal{C}(\mu_H)\right|^2$ through $O(\lambda^3)$ (with the ability to compute through $O(\lambda^4)$ if desired), this result is complete through $O(\lambda^3)$. One can use the relation between $\bar{\alpha}_L$ and $\alpha$ to re-write this expression in terms of the onshell coupling; however, we find it convenient to use the running coupling in our computations. 

We use \cref{factorized-form-Gamma-complete} to compute $1+\delta_R'(\mu_H)$ as presented in \cref{tab:decay_corrections}, evaluating $\tilde{F}(Z\bar{\alpha}_L,E,\mu_L)$ at $\mu_L=2\Em$, and using
\begin{equation}
    \bar{\alpha}_L = \alpha\qty(1+\frac{2}{3\pi} \alpha \log\qty(\frac{\mu_L}{\me}) + \frac{4}{9\pi^2} \alpha^2 \log^2\qty(\frac{\mu_L}{\me}) + \ldots )~.
\end{equation}
Unknown corrections stem from the finite contributions to $f^{(2,0)}$ and $f^{(3,2)}$ which cannot be inferred from the general arguments given above. 

Although not used explicitly in this work, it is useful to note that if one were to use an expansion in onshell $\alpha$ rather than $\bar{\alpha}_L$, then higher order terms must be kept to retain a complete $O(\lambda^3)$ counting. 
\begin{equation}
    \begin{split}
    \label{factorized-form-Gamma-complete-onshell}
    \Gamma = |\mathcal{C}(\mu_H)|^2 \left| \frac{\mathcal{C}(\mu_L)}{\mathcal{C}(\mu_H)}\right|^2 \int \dd \Phi_e \tilde{F}(Z \alpha ,E,\mu_L)  \bigg[1 + \delta^{(1,0)}(E) &+  \delta^{(2,1)}(E)  + \delta^{(3,2)}(E) \\
    &+ \delta^{(2,0)}(E) + \delta^{(3,1)}(E) +  O(\lambda^4) \bigg]~,
    \end{split}
\end{equation}
where, through $O(\lambda^3)$, we have
\begin{align}
    \left\langle \delta^{(2,0)}(E)  \right\rangle_0 &= 
     \alpha^2  \logM \left(-n_e \over 6\pi^2\right)  \left\langle \hat{g}(E,\mu) + {11\over 4} 
    \right\rangle_0~,
    \nl
    \left\langle \delta^{(3,2)}(E) \right\rangle_0 &= 
    Z^2\alpha^3 \logM 
    \left\langle 
    {\pi \over 6} - {5\over 8\pi}
    - {2n_e\over 3\pi} \left( -L_E + {\pi^2 \over 3} - {11\over 4} \right)
    \right\rangle_0~,
    \nl
    \left\langle \delta^{(3,1)}(E) \right\rangle_0 &=
    Z\alpha^3 \logM^2 \left\langle {-n_e^2\over 9\pi} \right\rangle_0~,
\end{align}
which can be inferred from the above expressions after relating $\bar{\alpha}_L$ to $\alpha$. The angle brackets denote averages over tree-level phase space in the $m/E\rightarrow 0$ limit,
\begin{equation}
    \langle \ldots \rangle_0 = \frac{30}{E_m^5} \int_0^{E_m} E^2 (E_m-E)^2 \qty[ \ldots ] ~. 
\end{equation}
For completeness, we tabulate also the corrections to the quantity $M$ through $\order(\lambda^3)$, collecting terms at first, second and third order in onshell $\alpha$ as 
\begin{align}
    M = 1 + \Delta^{(1,0)}+ \Delta^{(2,1)} +  \Delta^{(2,0)} + \Delta^{(3,2)} +   \Delta^{(3,1)} + \ldots, 
\end{align}
with
\begin{align}
    \Delta^{(1,0)} &\to {\alpha \over 2\pi} \left\langle \hat{g}(E,\mu) + {11\over 4} \right\rangle_0~ \nl
    &= 
    {\alpha \over 2\pi} \left( 3 \log{\mu\over 2 E_m} + {217 \over 20} - {4\pi^2\over 3} \right) \,,
    \nl
    \Delta^{(2,1)} &\to Z\alpha^2 \left[ 
     {n_e \over 3}\logM -\left( {2n_e\over 3} + 1 \right) \left( \log{2\Em\over \mu} - {47\over 60} \right) + {8n_e\over 9} + {11\over 4} -\zeta_2\
    \right] \,,\nl 
     \Delta^{(2,0)} &\to \alpha^2 \logM \left( -n_e \over 6\pi^2\right) \left( {217\over 20} - {4\pi^2 \over 3} \right) 
    \,,
    \\
    \Delta^{(3,2)} &\to Z^2 \alpha^3 \logM \left( -{2n_e \over 3 } \right) 
    \left[ -\left(\log{2\Em\over \mu} -{47 \over 60}\right) + {\pi^2 \over 3} + {11\over 4} \right]
    \,, \nl 
     \Delta^{(3,1)} &\to
    Z\alpha^3 \logM^2 \left({-n_e^2\over 9\pi} \right) \,. \nonumber
\end{align}
where the arrow denotes the limit $\me/\Em \to 0$.

\section{Relating schemes from Towner and Hardy to this work \label{app:relating-schemes}}
The calculation done in this work has been performed in the $\overline{\rm MS}$ scheme in a point-like effective theory that uses the physical nuclei as external degrees of freedom. Past calculations of radiative corrections have been performed using models that include spectator protons with a relativistic scalar propagator \cite{Sirlin:1967zza,Sirlin:1986cc,Jaus:1986te} (we will refer to this as ``Sirlin's scheme'' in what follows). This leads to some differences that must be accounted for when comparing results. For example, the Sirlin $g$-function (which arises naturally in the Sirlin-scheme) is defined by, 
\begin{equation}\label{eq:sirling}
    \begin{split}
    g(\Ee,\Em,m) = 3 &\log \left(\frac{m_p}{\me}\right)+ \frac{1}{\beta}\tanh ^{-1}(\beta ) \left(2 \left(\beta ^2+1\right)-4 \tanh ^{-1}(\beta )+\frac{(\Em-\Ee)^2}{6 \Ee^2}\right)\\
    &+4 \left(\frac{\tanh ^{-1}(\beta )}{\beta }-1\right) \left(\log \left(\frac{2 (\Em-\Ee)}{\me}\right)+\frac{\Em-\Ee}{3 E}-\frac{3}{2}\right)-\frac{4 \text{Li}_2\left(\frac{2 \beta }{\beta +1}\right)}{\beta }-\frac{3}{4}~,
    \end{split}
\end{equation}
with $\beta= |\vb{p}|/\Ee$ the electron velocity. In the EFT, no proton mass appears, and one instead finds the combination $g(\Ee,\Em,m) + 3\log\qty(\mu/m_p) + \frac{11}{4}$ where the additive constant $11/4$ is specific to the ${\overline{\rm MS}}$ scheme. Finite pieces, like $+11/4$, must be properly accounted for when comparing schemes. 

Towner and Hardy define their total set of radiative corrections as 
\begin{equation}
    (1+{\rm RC}) = 
    (1+\delta_R')_{\rm TH} (1+\delta_{\rm NS}-\delta_C) (1+\Delta_R^V) ~,
\end{equation}
where we have added a subscript of TH to distinguish from our $(1+\delta_R'(\mu_H))$. Of particular interest in the present context is their ``outer correction'', as defined in Eq.~(37) of \cite{Hardy:2020qwl} motivated by the factorization ansatz from Eq.~(15) of \cite{Czarnecki:2004cw}
\begin{equation}
    (1+\delta_R')_{\rm TH}=  \qty{1+ \frac{\alpha}{2\pi} \qty(\bar{g}-3\log\qty(\frac{m_p}{2\Em}))}\times \qty{L(2\Em, m_p) + \frac{\alpha}{2\pi}\qty( \delta_2^{({\rm TH})}+\delta_3^{({\rm TH})})}   ~, 
\end{equation}
where $\tfrac{\alpha}{2\pi} \bar{g}$ is taken from Table V of \cite{Towner:2007np}. It is convenient to define 
\begin{equation}
    \big[ \delta^{(1,0)} \big]_{\rm TH} \equiv \frac{\alpha}{2\pi} \qty(\bar{g}-3\log\qty(\frac{m_p}{2\Em}) + \frac{11}{4}) ~, 
\end{equation}
which involves only Towner and Hardy inputs, and is closely related\footnote{The numerical size of the difference ranges from $6\times 10^{-6}$ for $^{10}{\rm C}$ to  $1\times 10^{-4}$ for $^{54}{\rm Co}$. It stems from logarithmic terms in the Fermi function, $(\mu_H/\mu_L)^{2(\eta-1)} \sim 1 + Z^2\alpha^2\log(\mu_H/\mu_L)$, that are included in $R(\mu_H,\mu_L)$ in our analysis. } 
to our $\langle \delta^{(1,0}(E) \rangle_{\rm F}$. The term
\begin{equation}
    L(2\Em,m_p) = 1.026725\qty(1- \frac{2}{3\pi} \hat{\alpha}(\mu=\me) \log \frac{2\Em}{\me})^{9/4}~,
\end{equation}
is analogous to $R(\mu_H,\mu_L)$ discussed above, but containing only terms of $O(\alpha^nL^n)$.  The coupling with a hat is related to $\alpha$ by $\hat{\alpha}^{-1}(\mu=\me)=\alpha^{-1}+1/(6\pi)$ \cite{Czarnecki:2004cw}; following Towner and Hardy we take $\hat{\alpha}(\me) = 1/137.089$. Parts of $ \frac{\alpha}{2\pi} \delta_2^{\rm (TH)}$ (taken from Table V of \cite{Towner:2007np})  contain logarithms involving the nuclear scale $\Lambda$. Let us define $\delta^{(2,1)}_{\rm NS,TH}$ via, 
\begin{equation}
    \frac{\alpha}{2\pi} \delta_2^{({\rm TH})}= Z\alpha^2 \log\qty(\frac{\Lambda}{2\Em}) + \left\langle  \delta^{(2,1)}(E)\right\rangle_F + \delta^{(2,1)}_{\rm NS,TH}  ~,
\end{equation}
where $\delta^{(2,1)}_{\rm NS,TH}$ is an energy-independent constant related to the model calculation of Sirlin and Zucchini \cite{Sirlin:1986cc,Sirlin:1986hpu}. We have chosen the notation $\delta^{(2,1)}_{\rm NS,TH}$ because it may be interpreted as the short-distance region of Towner and Hardy's calculation \cite{Towner:2007np} and, therefore, as their estimate for the $Z\alpha^2$ correction to $|\mathcal{C}(\mu)|^2$. 

We may decompose $\delta_3$ using formulae from Sirlin and Zucchini \cite{Sirlin:1986cc,Sirlin:1986hpu}, which are given in terms of an energy dependent correction. Averaging
over phase space, and isolating logarithms of $\Lambda$ and $m$ we find,  
\begin{equation}
   \!\!\!\! \frac{\alpha}{2\pi} \delta_3^{({\rm SZ})} = Z^2 \alpha^3 \qty[ c_1  \log^2\qty(\frac{\Lambda}{2\Em})
    + c_2  \log\qty(\frac{2\Em}{\me})\log\qty(\frac{\Lambda}{2\Em}) + c_3 \log\qty(\frac{\Lambda}{2\Em})  + 
    c_4 \log\qty(\frac{2\Em}{\me}) + c_5] ~,
\end{equation}
where ``SZ'' stands for Sirlin-Zucchini, $c_5$ is $O(\lambda^4)$ in our counting and can be neglected, whereas $c_1=2/(3\pi)$, $c_2=4/(3\pi)$, 
\begin{align}
    c_3&=\frac{6\pi^2-67 - 12\log(10)}{18\pi}  ~,\nl
    c_4&=-\frac{20 \pi ^2+48 \gamma_{\rm E} +24 \log (10)-177}{36 \pi } - c_2\left\langle \log(\tfrac{\Ee}{\Em}) \right\rangle_0 ~.  
\end{align}
We have used $(2\pi)^{-1}(\frac{5}{2}-\frac{2 \pi ^2}{3})\approx -0.649$ for Sirlin and Zucchini's $h$ from \cite{Sirlin:1986cc}. We have also accounted for the updates described in the footnotes of \cite{Sirlin:1986hpu} which include adjusting to the 1986 Jaus-Rasche updated logarithmic coefficient $a=\pi/3-3/(2\pi)$ \cite{Jaus:1986te} (which is erroneous \cite{Borah:2024ghn}), and using $\log(\Ee/\me)$ rather than $\log(\Em/\me)$ for the term multiplying $h$ as given in \cite{Sirlin:1986cc}. 

Towner and Hardy's tabulation \cite{Towner:2007np} disagrees with the tabulation found in Sirlin \cite{Sirlin:1986hpu}. We have calculated $\frac{\alpha}{2\pi} \delta_3$ using the above formula and find that if we set $c_4=c_5=0$ we reproduce all of Towner and Hardy's tabulated results \cite{Towner:2007np} except for $^{50}{\rm Mn} \rightarrow \! ~^{50}{\rm Co}$ for which we obtain $6.3\times10^{-4}$ as compared to Towner and Hardy's $6.2\times10^{-4}$. We therefore conclude that the tabulated values in \cite{Towner:2007np}, $\frac{\alpha}{2\pi} \delta_3^{({\rm TH})}$, are equivalent to,
\begin{equation}
    \frac{\alpha}{2\pi} \delta_3^{({\rm TH})} \leftrightarrow Z^2 \alpha^3 \qty[ c_1  \log^2\qty(\frac{\Lambda}{2\Em})
    + c_2  \log\qty(\frac{2\Em}{\me})\log\qty(\frac{\Lambda}{2\Em}) + c_3 \log\qty(\frac{\Lambda}{2\Em}) ] ~,
\end{equation}
which may be interpreted as part of the renormalization group flow. 

Without using any new inputs, it is then natural to re-organize Towner and Hardy's results as 
\begin{align}
    \label{short-distance}
   (1+\delta_{\rm NS}-\delta_C) (1+\Delta_R^V) &\rightarrow  \underbrace{L(\Lambda,m_p)\bigg(1+\delta_{\rm NS} + \delta_{\rm NS,TH}^{(2,1)}-\delta_C\bigg) \bigg(1+\Delta_R^V - \frac{11}{8\pi} \alpha\bigg) }_{\rm Short-distance}~,\\
    (1+\delta_R')_{\rm TH} &\rightarrow \underbrace{\qty{1+ \big[ \delta^{(1,0)} \big]_{\rm TH} + \big\langle  \delta^{(2,1)}(E)\big\rangle_F  } \mathcal{R}(2\Em,\Lambda)}_{\rm Long-distance} ~,
    \label{long-distance}
\end{align}
where we have split up $L(2\Em,m_p)= L(2\Em,\Lambda)L(\Lambda,m_p)$ and introduce the analog of our RG factor, 
\begin{equation}
    \label{script-R-def}
    \mathcal{R}(2\Em,\Lambda)= L(2\Em, \Lambda) + Z\alpha^2 \log\qty(\tfrac{\Lambda}{2\Em})+ \frac{\alpha}{2\pi} \delta_3^{({\rm TH})} ~,
\end{equation}
which now contains the $Z\alpha^2 L $ term and Towner and Hardy's estimate for the $Z^2\alpha^3$ logarithmic terms taken explicitly from Table V of \cite{Towner:2007np}. The quantity $\left\langle  \delta^{(2,1)}(E)\right\rangle_F$ is expanded in terms of the onshell coupling $\alpha$, and evaluated at $\mu_L= 2\Em$. The RG-running factor from $\Lambda$ to $2\Em$ is computed following \cite{Czarnecki:2004cw} for self-consistency with the Towner and Hardy result,\!\footnote{The numerical coefficient $1.01882$ is obtained using Eq.~(56) of Ref.~\cite{Czarnecki:2004cw}, and includes the running from the electron mass threshold up to the muon mass threshold in their given formalism.}
\begin{equation}
    \label{L-factor-calc}
    L(2\Em,\Lambda)= 1.01882 \qty(1- \frac{2}{3\pi} \hat{\alpha}(\me) \log \frac{2\Em}{\me})^{9/4}   \qty(\frac{1}{1-\frac{22}{9\pi}\hat{\alpha}(m_\mu)\log(\Lambda/m_\mu)})^{27/44}~,
\end{equation}
where $\hat{\alpha}(m_\mu)= 1/135.7896$. While \cref{L-factor-calc} matches \cite{Czarnecki:2004cw} and includes constituent quarks and a dynamical muon, we have checked that the same numerical value is obtained, up to corrections of $10^{-5}$ if one includes only dynamical electrons in the theory. The result, and therefore our comparison to Towner and Hardy, is insensitive to the detailed implementation.

The replacement rules defined above do not introduce any new ingredients to Towner and Hardy's analysis. Instead, they simply re-arrange existing inputs in accord with expectations from the factorization theorems discussed in the main text. We, therefore, refer to the resulting change as a ``convention shift'' defined via, 
\begin{equation}
    \text{Convention shift}~ = ~
    \frac{(\text{Long-distance})\times (\text{Short-distance})}{ (1+\delta_R')_{\rm TH} (1+\delta_{\rm NS} -\delta_C)(1+\Delta_R^V)}~,
\end{equation}
where the long- and short-distance pieces are defined by \cref{short-distance,long-distance}. The convention shift receives a contribution of $\sim 0.85\times10^{-4}$ from induced cross terms between $\tfrac{11}{8\pi} \alpha$ and $\Delta_R^V$ or $L(2 E_m,m_p)$. Other contributions are formally $O(\lambda^4)$ in our counting if $\log(m_p/\Lambda)$ is counted as $O(1)$, such as the cross term of $(Z\alpha^2 L)\times  ( \tfrac{3\alpha}{2\pi} \log\tfrac{m_p}{\Lambda})$ with the $\log(m_p)$ term coming from $L(\Lambda,m_p)$.  

Next, we measure the impact of our new inputs by taking the ratio of our model-independent calculation of the long-distance corrections to that defined using Towner and Hardy inputs, 
\begin{equation}
    \text{Long-distance shift}~ = ~
    \frac{1+\delta_{R}'(\mu_H=\Lambda)}{(\text{Long-distance})}~. 
\end{equation}
This comparison accounts for new information provided by our analysis beyond the structure of the amplitude's factorization. For example, the coefficient of the $\log(\Lambda/ 2\Em)$ term in \cref{script-R-def} should read $(\frac{\pi}{6}-\frac{3}{\pi})$ \cite{Borah:2024ghn}.  Similarly the $\frac{13}{6\pi} Z \alpha^2 L^2$ term in \cref{R-expl-ms-bar} is a new effect not included in past analyses. 

\pagebreak 

\section{Structure dependent effects \label{app:2008-reproduce} }
\begin{table}[ht]
\centering
\begin{tabular}{|c|c|c|c||c| c |} \hline 
    Beta Emitter & $F_{0}(Z,E)$ & $F_{\rm BJ}(Z,E)$ & $F_{\rm BJ}(Z,E)\times C(Z,E)$  & {\rm Towner \& Hardy} & 
    Error estimate [$10^{-4}]$\\ \hline \hline
    $\tensor[^{10}]{\text{C}}{}$ & 1.467 & 1.467 & 1.467 & 1.468 & 0.1 \\
    $\tensor[^{14}]{\text{O}}{}$ & 1.286 & 1.285 & 1.285 & 1.286 & 0.1 \\
    $\tensor[^{18}]{\text{Ne}}{}$ & 1.204 & 1.204 & 1.203 & 1.204 & 0.1 \\
    $\tensor[^{22}]{\text{Mg}}{}$ & 1.122 & 1.121 & 1.121 & 1.122 & 0.1 \\
    $\tensor[^{26}]{\text{Si}}{}$ & 1.056 & 1.055 & 1.054 & 1.055 & 0.1 \\
    $\tensor[^{30}]{\text{S}}{}$ & 1.007 & 1.005 & 1.005 & 1.005 & 0.2 \\
    $\tensor[^{34}]{\text{Ar}}{}$ & 0.966 & 0.963 & 0.962 & 0.963 & 0.3 \\
    $\tensor[^{38}]{\text{Ca}}{}$ & 0.932 & 0.929 & 0.928 & 0.929 & 0.3 \\
    $\tensor[^{42}]{\text{Ti}}{}$ & 0.911 & 0.907 & 0.906 & 0.906 & 0.5 \\
    \hline \hline
    $\tensor[^{26}]{\text{Al}}{^m}$ & 1.110 & 1.109 & 1.109 & 1.110 & 0.1 \\
    $\tensor[^{34}]{\text{Cl}}{}$ & 1.004 & 1.002 & 1.002 & 1.002 & 0.2 \\
    $\tensor[^{38}]{\text{K}}{^m}$ & 0.967 & 0.964 & 0.963 & 0.964 & 0.3 \\
    $\tensor[^{42}]{\text{Sc}}{}$ & 0.943 & 0.940 & 0.939 & 0.939 & 0.4 \\
    $\tensor[^{46}]{\text{V}}{}$ & 0.909 & 0.905 & 0.903 & 0.903 & 0.6 \\
    $\tensor[^{50}]{\text{Mn}}{}$ & 0.880 & 0.875 & 0.873 & 0.873 & 0.7 \\
    $\tensor[^{54}]{\text{Co}}{}$ & 0.854 & 0.847 & 0.844 & 0.844 & 1.0 \\
    $\tensor[^{62}]{\text{Ga}}{}$ & 0.820 & 0.810 & 0.806 & 0.805 & 1.5 \\
    $\tensor[^{66}]{\text{As}}{}$ & 0.808 & 0.797 & 0.792 & 0.791 & 1.7 \\
    $\tensor[^{70}]{\text{Br}}{}$ & 0.796 & 0.783 & 0.777 & 0.776 & 2.0 \\
    $\tensor[^{74}]{\text{Rb}}{}$ & 0.785 & 0.770 & 0.763 & 0.761 & 2.4 \\ \hline
\end{tabular}
\caption{ Calculation of 
    $\frac{\alpha}{2\pi} \overline{g}(E_{\me})$, in units of percent, with the conventions and inputs used for Table V of \cite{Towner:2007np} (taken from \cite{Hardy:2004id}). The first column uses the point-like Fermi function as defined in \cref{eq:F0-def}. The second column uses instead $F_{\rm BJ}(Z,E)$, the Behrens and J\"anecke Fermi function defined in \cref{eq:FBJ-def}. The third column uses the product of $F_{\rm BJ}(Z,E)$ and the shape function $C(Z,E)$ as given in Ref.~\cite{Hayen:2017pwg}. We find that agreement between our implementation and that tabulated in Towner and Hardy is, in general, improved after structure-dependent corrections are included, with this improvement more pronounced for heavier nuclei. 
    The radiative corrections to higher dimensional operators are not guaranteed to be captured by multiplication with the Sirlin-$g$ function. We therefore use the difference between the Towner and Hardy values and our point-like calculation as an estimate of the uncertainty due to the standard factorization ansatz used in the beta decay literature; if the difference is zero, we take the largest fluctuation in the row. 
    \label{tab:Delta_points}}
\end{table}

As a check of our averaging procedure, we have taken all of the numerical inputs listed in Towner and Hardy's 2004 review \cite{Hardy:2004id}, and attempted to reproduce the value of $\bar{g}$ quoted in their 2007 article \cite{Towner:2007np}.  This both validates that the averaging procedure used here is consistent with Towner and Hardy's analysis, and allows us to estimate errors associated with the factorization ansatz for one-loop corrections typically used in the literature. As discussed in \cref{app:one-loop-comment}, this ansatz turns out to be exact at leading-power but is not guaranteed when including energy-dependent terms proportional to the nuclear radius which correspond to the inclusion of higher dimensional operators in the effective theory.

Towner and Hardy compute their phase space averages with numerical solutions of the Dirac equation \cite{Hardy:2004id}. In our comparison we do not solve the Dirac equation numerically and instead use analytic formulae taken from \cite{Hayen:2017pwg}. The first step is to use the conventional point-like Fermi function,
\begin{equation}\label{eq:F0-def}
    F_0(Z,\Ee) = 4 (2pR)^{2(\eta-1)} \e^{-\pi \xi} \frac{|\Gamma(\eta-\iu\xi)|^2}{|\Gamma(1+2\eta)|^2}~.
\end{equation}
The radius $R$ is defined by $R=(5/3)^{1/2} \langle r^2 \rangle^{1/2}$ and we use the same root-mean-square radii as above (these can be inferred from $\Lambda$ using $ \langle r^2 \rangle^{1/2} = \sqrt{6}/\Lambda$). As we show in \cref{tab:Delta_points}, and discuss below, this does a good job of replicating Towner and Hardy's calculations, with some discrepancies at the level of $\sim 10^{-4}$ for heavier nuclei. 

To investigate the origin of these discrepancies, we have tried three different methods for computing phase space averages. The first involves only the Fermi function $F_0(Z,E)$ as describe above. Second, we replace $F_0$ by the ``Behrens and J\"anecke'' \cite{Schopper:1969jkp} Fermi function as given in Eq.~(10) of \cite{Hayen:2017pwg},
\begin{equation}\label{eq:FBJ-def}
    F_{\rm BJ}(Z,\Ee) = F_0(Z,\Ee) \qty[ 1 + \frac{13}{15}Z\alpha \Ee R ] ~,
\end{equation}
which incorporates some nuclear structure effects. As a third approach we multiply $F_{\rm BJ}(Z,E)$ by the shape factor, $C(Z,E)$, as parameterized in Eqs.~(100) and (101) of \cite{Hayen:2017pwg}. 

The numerical results of these three averaging procedures are shown in \cref{tab:Delta_points}. The point-like Fermi function, $F_0(Z,E)$, fails to reproduce the full set of Towner and Hardy's values of $\frac{\alpha}{2\pi} \bar{g}$ with discrepancies of $\sim 10^{-4}$.  
Inclusion of structure dependent effects, parameterized with the analytic formulae discussed above, improves agreement to $\sim 10^{-5}$. We conclude that there is a  $\sim 10^{-4}$ uncertainty for heavier nuclei, due to the conventional factorization ansatz for radiative corrections.

\end{fmffile}

\bibliography{Z_alpha_squared}

\end{document}